\def\asca{{\it ASCA\/}}
\def\chandra{{\it Chandra\/}}

\def\hst{{\it {\it HST}\/}}

\def\rosat{{\it ROSAT\/}}

\def\aox{$\alpha_{\rm ox}$}
\def\vla{VLA~J123642.09+621331.4} 
\def\nicmos{NICMOS~J123651.74+621221.4} 

\def\ltsima{$\; \buildrel < \over \sim \;$}
\def\simlt{\lower.5ex\hbox{\ltsima}}
\def\gtsima{$\; \buildrel > \over \sim \;$}
\def\simgt{\lower.5ex\hbox{\gtsima}}
        
\documentclass[preprint]{aastex}

\begin{document}


\title{The Chandra Deep Survey of the Hubble Deep Field North Area. IV.
An Ultradeep Image of the HDF-N}


\author{W.N.~Brandt,$^1$ A.E.~Hornschemeier,$^{1}$ D.M.~Alexander,$^{1}$ G.P.~Garmire,$^1$ D.P.~Schneider,$^1$
P.S.~Broos,$^1$ L.K.~Townsley,$^1$
M.W.~Bautz,$^2$ E.D.~Feigelson,$^1$ and R.E.~Griffiths$^3$ 
%
%
%
%
}

\footnotetext[1]{Department of Astronomy \& Astrophysics, 525 Davey Laboratory, 
The Pennsylvania State University, University Park, PA 16802}

\footnotetext[2]{Massachusetts Institute of Technology, Center for Space Research, 
70 Vassar Street, Building 37, Cambridge, MA 02139}

\footnotetext[3]{Department of Physics, Carnegie Mellon University, Pittsburgh, PA 15213}









\begin{abstract}
We present results from a 479.7~ks \chandra\ exposure of the Hubble Deep Field
North (HDF-N) and its immediate vicinity. In this X-ray image, the deepest ever
reported with a 0.5--2.0~keV flux limit of $\approx 4.9\times 10^{-17}$~erg~cm$^{-2}$~s$^{-1}$, 
four new HDF-N \hbox{X-ray} sources are detected bringing the total number of such sources to 12. 
The new sources include two optically bright ($R=$~18.3--18.8), low-redshift ($z<0.15$)
galaxies, a Fanaroff-Riley~I radio galaxy, and an edge-on spiral hosting either a 
powerful starburst and/or a low-luminosity active galactic nucleus (AGN). Notably, 
X-ray emission has now been detected from {\it all\/} 
luminous galaxies ($M_{\rm V}<-18$) with $z<0.15$ 
known in the \hbox{HDF-N}. We have also detected the remarkable $\mu$Jy radio source \vla, 
which is located just outside the \hbox{HDF-N} and has a likely redshift of $z=4.424$. 
The observed X-ray emission supports the presence of an AGN in this object, and its 
X-ray-to-optical flux ratio (i.e., \aox) is consistent with what is seen for 
low-redshift AGN.

We have detected X-ray variability from two of the previously known HDF-N 
X-ray sources, and spectral fitting shows clear evidence for X-ray 
absorption in the brightest X-ray source in the HDF-N, 
a $z=0.960$ broad-line AGN with associated Mg~{\sc ii} absorption. Stacking analyses of
optically bright HDF-N galaxies not individually detected in X-rays have 
provided estimates of their average X-ray fluxes, and we find that the
X-ray luminosities of ``normal'' spirals at $z\approx 0.5$ are not more than a factor of
$\approx 2$ larger (per unit $B$-band luminosity) than 
those of spirals in the local Universe ($z<0.01$).
This constrains models for the evolution of low-mass X-ray binary populations
in galaxies in response to the declining cosmic star-formation rate. 
Monte-Carlo simulations support the validity of the stacking analyses
and show that the \chandra\ Advanced CCD Imaging Spectrometer (ACIS) performs
source detection well even with effective exposure times of $\approx 8$~Ms. 
\end{abstract}


\keywords{
diffuse radiation~--
surveys~--
cosmology: observations~--
galaxies: active~--
X-rays: galaxies~--
X-rays: general.}


\section{Introduction \label{intro}}

We are in the process of performing a deep X-ray survey ($\approx 1$~Ms) 
of the Hubble Deep Field North (HDF-N; Williams et~al. 1996, hereafter W96; 
Ferguson, Dickinson, \& Williams 2000) and its environs with the 
{\it Chandra X-ray Observatory\/} (hereafter \chandra; Weisskopf et~al. 2000). 
This field was chosen because of the excellent radio, submillimeter, 
infrared, and optical studies already completed (see Livio, Fall, \& Madau 1998
and Ferguson et~al. 2000 for reviews). The HDF-N itself is the most 
intensively studied extragalactic patch of sky, and it is
one of only two such regions (the other being the Hubble Deep Field
South) with high-resolution imaging down to 28--29$^{\rm th}$ 
magnitude in several optical bands. By centering our survey on the HDF-N, 
we ensure that the best possible data will be available for follow-up of the 
faintest X-ray sources ever detected. 

In Hornschemeier et~al. (2000, hereafter H00), we presented the first 
results for the HDF-N based on 164.4~ks of \chandra\ exposure. These
results were extended to 221.9~ks over an area much larger than the
HDF-N in Hornschemeier et~al. (2001, hereafter H01)
and G.P. Garmire et~al., in preparation (hereafter G01); eight 
sources in the HDF-N itself were reported. The total exposure on 
the HDF-N field has recently been more than doubled to 479.7~ks;
these new data have allowed the detection of several additional HDF-N 
sources and substantially improved the X-ray constraints on the 
previously detected sources. Here we present the additional sources 
as well as improved constraints on sources in the HDF-N itself 
and on one source in its immediate vicinity. We also use
stacking analyses to probe the average X-ray emission properties of
HDF-N galaxies that are not individually detected in the X-ray band. 

The Galactic column density along this line of sight
is $(1.6\pm 0.4)\times 10^{20}$~cm$^{-2}$ (Stark et~al. 1992).
$H_0=70$~km~s$^{-1}$ Mpc$^{-1}$ and $q_0=0.1$ are adopted throughout this paper.
Coordinates throughout this paper are J2000.


\section{Chandra ACIS Observations and Analysis\label{ACISobs}}

\subsection{Observation Details and Image Creation}

The field containing the HDF-N was observed with the \chandra\ Advanced 
CCD Imaging Spectrometer (ACIS; G.P. Garmire et~al., in preparation) 
for a total exposure time of 257.9~ks on 2000 November 21--23 
(167.0~ks; observation ID 1671) and 2000 November 24--25 
(90.9~ks; observation ID 2344). The background was stable during the
first observation, but there was significant background
flaring due to ``space weather'' during $\approx 30$~ks of the
second observation. During this $\approx 30$~ks the background
was $\approx 2$ times higher, but this has had little impact on 
the analysis or results presented here. The HDF-N was placed near the aim point 
for the ACIS-I array on CCD I3 during both observations. These data were 
added to the 221.9~ks of data presented in H01; the entire HDF-N was 
covered for all observations and was kept away from gaps between the CCDs. 

Several improvements over the data reduction techniques of H01 have been made. 
The most recently processed data from the \chandra\ X-ray Center were used, 
and the standard $0.5^{\prime\prime}$ pixel randomization was 
removed.\footnote{See http://asc.harvard.edu/cal/Hrma/hrma/misc/oac/dd\_psf/dd\_randomiz.html.}
\chandra\ Interactive Analysis of Observations (CIAO) Version~2.0 
tools were used whenever possible in place of some of the custom software used 
for H01.\footnote{See http://asc.harvard.edu/ciao/.} 
All data were corrected for the radiation damage sustained by the CCDs during 
the first few months of \chandra\ operations using the procedure of 
Townsley et~al. (2000). This procedure partially corrects for the positionally dependent 
grade distribution due to inefficient charge transfer in the radiation-damaged
CCDs. It also partially corrects for quantum efficiency losses, which are most 
significant in non-corrected data at $-110^{\circ}$~C (see Townsley et~al. 2000 
and H01 for discussion of the remaining small quantum efficiency 
losses incurred).
The absolute astrometry was determined following 
the method in \S3.1 of H01; absolute X-ray source positions in the 
HDF-N itself are accurate to $0.4^{\prime\prime}$ or better. 

We created images from 0.5--8.0~keV (full band), 0.5--2.0~keV (soft band),
2--8~keV (hard band) and 4--8~keV (ultrahard band) using the two event 
grade sets defined in Table~1; hereafter these will be called the
``standard \asca\ grade set'' and the ``restricted ACIS grade set.'' 
The use of the restricted ACIS grade set improves our ability to 
detect faint sources in some cases, as discussed in Brandt et~al. (2000) 
and T.~Miyaji et~al., in preparation. The basic reason for this improvement is 
that the restricted ACIS grade set rejects a significantly higher fraction 
of background events than source events (after correction for the 
radiation damage). 
With the restricted ACIS grade set, the average soft-band and hard-band 
background levels are reduced by 36\% and 28\% relative to the levels
with the standard \asca\ grade set (see Table~1). 
For comparison, the bright source CXOHDFN~J123646.3+621404 (see below) 
loses 12\% of its soft-band counts and 14\% of its hard-band counts
when the restricted ACIS grade set is used instead of the standard
\asca\ grade set. 

\subsection{Source Detection and Detection Limits}

Source detection was performed with {\sc wavdetect} (Freeman et~al. 2001). 
Our criterion for source detection is that a source must be
found with a false-positive probability threshold of $1\times 10^{-7}$ in 
at least one of the four bands using either the standard \asca\ or restricted ACIS 
grade sets. The 12 sources found in this manner are listed in Table~2.
All photometry in Table~2 is for the standard \asca\ grade set. In Table~2 
we also report results for one \chandra\ source 
located just outside the HDF-N, CXOHDFN~J123642.0+621331. 
Figure~1 shows the detected sources in the full, soft and hard bands
(the ultrahard-band image is not shown since only two sources are detected).
Some of the sources outside the HDF-N that are not labeled in Figure~1
are discussed in H01, while others will be discussed in a future publication. 
Conservatively treating the eight images searched as entirely independent, 
$\simlt 0.15$ false sources are expected statistically. All sources have 
been manually inspected to be certain that they are not produced or affected
by ``cosmic ray afterglows'' (\chandra\ X-ray Center 2000, private communication). 
Finally, images made with other grade set choices (e.g., only ACIS grade~0 
events) were searched; no additional sources were detected. 

{\sc wavdetect} was also run in the various bands to search for 
lower-significance counterparts (matching to within $0.5^{\prime\prime}$) 
of the highly significant sources already detected at the $1\times 10^{-7}$ 
level in at least one of the four bands; in these runs we used 
probability thresholds of $1\times 10^{-6}$ and $1\times 10^{-5}$. Since 
the spatial-matching requirement greatly reduces the number of pixels being 
searched, the statistically expected number of false cross-band matches 
obtained in this manner is very small. The only two cases where new 
cross-band matches were found with probability thresholds of 
$1\times 10^{-6}$ or $1\times 10^{-5}$ are noted in Table~2. 

The nature of the multiwavelength counterparts to our new X-ray sources
supports the validity of the X-ray detections (see \S3.1 for details on
the multiwavelength counterparts). For example, four 
of the five new X-ray sources align with radio sources
(see Table~3), extending the trend established for this field at 
brighter X-ray fluxes (H00; H01). Given the precise X-ray and 
radio positions, the a priori probability 
of such an alignment is small ($<10^{-7}$). 
Similarly, the probability of alignment with optically bright, 
low-redshift galaxies is small (compare with \S4.1 of H01). 
The one new X-ray source in \S3.1 that is not a radio source 
aligns with an optically bright galaxy. 

The ``effective exposure time'' per source, as derived from our exposure 
map, ranges from 444--464~ks (see H01 for 
discussion). Even with these long exposure times,
the \chandra\ ACIS is entirely photon limited for point-source detection
near the aim point. For a power-law model with photon index $\Gamma=1.4$ 
and the Galactic column density, our $\approx 5$ count
detection limit corresponds to soft-band and hard-band flux limits of  
\hbox{$\approx 4.9\times 10^{-17}$~erg~cm$^{-2}$~s$^{-1}$} and 
\hbox{$\approx 2.3\times 10^{-16}$~erg~cm$^{-2}$~s$^{-1}$}, respectively. 

\subsection{Variability Testing}

All sources have been checked for count-rate variability using 
Kolmogorov-Smirnov (K-S) tests, and the results are reported in 
Table~2. Such testing is statistically valid even when the number 
of counts is small, although clearly the sensitivity of the K-S 
test is reduced when only a few counts are available. Use of 
the K-S test avoids a posteriori selection problems as
discussed by Press \& Schechter (1974). The testing is always
performed in the band where a source has its highest signal-to-noise 
ratio, and the standard \asca\ grades were used in this analysis.
We have used $3^{\prime\prime}\times 3^{\prime\prime}$ apertures 
for count extraction; these are appropriate since the HDF-N was
kept close to the aim point for all observations. Gaps between 
observations were accounted for in the testing. 

The two sources showing statistically significant evidence for 
variability are discussed in \S3.2. 


\section{Results\label{Results}}

\subsection{Newly Detected Sources}

Figure~2 shows the \chandra\ sources detected thus far overlaid on 
the W96 HDF-N image, and Table~3 gives the identification information for
these sources. Below we describe the properties of the newly 
detected sources. 

{\it CXOHDFN~J123642.0+621331:}
This X-ray source lies just outside the HDF-N itself and is positionally coincident 
with the remarkable $\mu$Jy radio source \vla\ (e.g., Richards et~al. 1998; Garrett et~al. 2001). 
Detailed studies of this source indicate a redshift of $z=4.424$
(Waddington et~al. 1999; R.A. Windhorst 2001, private communication; but also 
see \S2.2 of Barger, Cowie, \& Richards 2000); we shall adopt this redshift throughout 
the rest of the discussion below as the other redshift possibilities appear less likely. 
This source is thought to be powered by either a dust-enshrouded AGN or an ultraluminous 
starburst. If the redshift is correct, then this is the most distant X-ray source detected 
thus far in our \chandra\ field and the ninth most distant cosmic object detected
in X-rays (after seven quasars and one gamma-ray burst). The 
0.5--2.0~keV observed-frame bandpass in which \vla\ is most clearly detected 
corresponds to a rest-frame bandpass of 2.7--10.8~keV. The implied 
luminosity in this bandpass of $\approx 2\times 10^{43}$~erg~s$^{-1}$
strongly supports the presence of an AGN.

In Figure~3 we compare the AB$_{1450}$ magnitude (AB$_{1450}\approx 25.2$ 
from Figure~3 of Waddington et~al. 1999) and 0.5--2.0~keV flux of \vla\ with 
those of \hbox{$z>4$} AGN observed in X-rays (see Kaspi, Brandt, \& Schneider 2000
and Brandt et~al. 2001). \vla\ is both the optically faintest and 
X-ray faintest $z>4$ AGN detected in X-rays; it provides the 
best constraints to date upon the X-ray properties of moderate-luminosity 
$z>4$ AGN. In fact, to our knowledge \vla\ has the lowest optical luminosity
of all known $z>4$ AGN (compare with Stern et~al. 2000). 

We have estimated \aox, the slope of a nominal power law between 
2500~\AA\ and 2~keV in the rest frame. A $\Gamma=2$ power law was 
adopted for the X-ray continuum (following Reeves \& Turner 2000), and 
an optical power-law slope of $\alpha_{\rm o}=-0.79$ 
(see Schneider, Schmidt, \& Gunn 1991 and Fan et~al. 2001) was used
to calculate the 2500~\AA\ flux density from the AB$_{1450}$ 
magnitude. We find $\alpha_{\rm ox}=-1.46$. This value of \aox\ is 
in good agreement with \aox\ values found for low-redshift AGN of 
comparable X-ray luminosity (e.g., Avni \& Tananbaum 1986; 
Brandt, Laor, \& Wills 2000, hereafter BLW).
There is no evidence that the X-rays are strongly suppressed by internal 
absorption, although given that we are only sensitive to hard X-rays
in the rest frame the intrinsic $N_{\rm H}$ is only constrained to 
be $\simlt 5\times 10^{23}$~cm$^{-2}$. 


In contrast to the results from optically and radio selected AGN surveys, 
\rosat\ surveys do not show a significant decrease in the space density 
of high-redshift, high-luminosity AGN (see \S5 of Miyaji, Hasinger, \& Schmidt 2000). 
Our X-ray detection of the moderate-luminosity AGN \vla\ at $z>4$ 
should ultimately be useful for pinning down the X-ray 
luminosity function (XLF) of AGN when combined with the results from other
deep X-ray surveys. At present it is difficult to assess the relation of \vla\ to 
the XLF because its X-ray luminosity is $\approx 10$ times lower than those used 
for all high-redshift XLF studies to date. 

{\it CXOHDFN~J123644.3+621132:}
This X-ray source is coincident with a Fanaroff-Riley~I (FR~I) 
radio galaxy at $z=1.050$ that is one of the brightest radio sources 
in our field (its 1.4~GHz flux density is 1.29~mJy; Richards et~al. 1998; 
Richards 2000). Its radio structure is extended over 
$\approx 40^{\prime\prime}$, but we only detect the core in X-rays
(see Figure~4). The host galaxy is a large elliptical and is a Very
Red Object (VRO) with ${\cal R}-K_{\rm s}=5.10$ (Hogg et~al. 2000). 

Using stacking techniques, H01 found evidence for two X-ray emission
classes of VROs: (1)~X-ray luminous VROs with hard X-ray spectra and
(2)~lower X-ray luminosity VROs with softer X-ray spectra. 
CXOHDFN~J123644.3+621132 is only detected in the soft band and
appears to be a member of the second class that has now been individually 
detected (previously the X-ray emission from the second class was only 
detectable when many objects were stacked). The X-ray emission 
is probably associated with the AGN or the hot interstellar medium 
and X-ray binaries of the host galaxy. The observed band ratio (see 
Table~2) cannot discriminate between these possibilities, and the 
X-ray luminosity is not inconsistent with an origin in the
hot gas and X-ray binaries of this large elliptical
(e.g., see Figure~2 of Matsushita, Ohashi, \& Makishima 2000). 

FR~I sources are often located in clusters of galaxies, but no evidence 
is found for diffuse X-ray emission around CXOHDFN~J123644.3+621132. 
At $z=1.050$, the angular diameter of the core of a cluster would be
$\approx 1^{\prime}$ (corresponding to $\approx 0.5$~Mpc). Using an annulus 
with an inner radius of $1.5^{\prime\prime}$ and an outer radius of 
$34^{\prime\prime}$, we find a $3\sigma$ upper limit on the number of
full-band counts of $<120$ (we also excluded X-ray emission from 
CXOHDFN~J123641.7+621131 using an exclusion cell with a radius
of $1.5^{\prime\prime}$). For a Raymond-Smith plasma with $kT=4$~keV 
and one-third solar abundances, the full-band luminosity limit is 
$<8.4\times 10^{42}$~erg~s$^{-1}$. 
No hint of X-ray emission from the other $z\approx 1.050$ objects near 
CXOHDFN~J123644.3+621132 is seen (see \S6.1 of Richards et~al. 1998). 

{\it CXOHDFN~J123648.3+621426:}
This source is coincident with a bright $z=0.139$ elliptical located
at the edge of the HDF-N (see Figure~5). This galaxy 
has also been detected at radio and mid-infrared 
wavelengths (Richards et~al. 1998; Aussel et~al. 1999). 
The mid-infrared colors from {\it ISO\/} are consistent with the re-emission
of starlight by interstellar grains as opposed to the light from cool
stars (Rowan-Robinson et~al. 1997). Given the low X-ray luminosity of 
this source and the weak constraint on its band ratio of $<0.80$ 
(see Table~2; this band ratio constraint corresponds to a photon
index constraint of $\Gamma>1$ for a power-law model with the Galactic column
density), the observed X-ray emission can be plausibly explained by 
emission from a hot interstellar medium, perhaps combined with emission 
from X-ray binaries (e.g., Matsushita et~al. 2000 and references therein). 

{\it CXOHDFN~J123649.5+621345:}
This source is located $\approx 1.33^{\prime\prime}$ from the center of one 
of the optically brightest galaxies in the HDF-N, an S0 or elliptical at
$z=0.089$ (see Figure~5); the corresponding projected physical offset is 
$\approx 2.6$~kpc. Although the photon statistics are limited, the fact
that this offset is $\simgt 3$ times larger than the offsets for the other
sources (see Table~3) argues that the X-ray source is indeed off-nuclear. 
The 0.5--8.0~keV luminosity of $\approx 2\times 10^{39}$~erg~s$^{-1}$
is comparable to that of the most luminous X-ray binaries in 
the Local Group, and an X-ray binary nature for this source would 
naturally explain its off-nuclear location. We are not able to constrain
the band ratio for this source since it is only detected in the 
full band. 

{\it CXOHDFN~J123649.7+621312:}
The X-ray source is coincident with the nucleus of an edge-on spiral at 
$z=0.475$ (see Figure~5) that is also a steep-spectrum radio source 
(Richards et~al. 1998). The radio emission is thought to be predominantly 
due to starburst activity (see Chapter~4 of Richards 1999). The soft-band 
X-ray luminosity of $\approx 5\times 10^{40}$~erg~s$^{-1}$ is $\approx 5$ 
times that of the prototype starburst galaxy M82 (e.g., Griffiths et~al. 2000). 
Given the X-ray luminosity and X-ray-to-optical flux ratio, the observed
emission could be created by a powerful starburst and/or a low-luminosity 
active nucleus (e.g., Moran, Halpern, \& Helfand 1996; 
Boller et~al. 1998; Ho et~al. 2001). 



\subsection{Further Results on Previously Detected Sources}

The other eight \chandra\ sources detected in the HDF-N have already been described 
in H00 and \S6 of H01; all of these previously reported sources are still 
detected in our improved analysis here. Below we report on some new results 
for these sources.

{\it CXOHDFN~J123639.5+621230:}
This source is associated with a $z=3.479$ broad-line AGN (see H01). 
The K-S test indicates highly significant count-rate variability 
in the full band; a constant count rate is rejected with $>99.9$\% 
confidence (see Table~2). The variability is also apparent in Figure~6. 
During the first four observations, 15 source plus background
counts were obtained in 221.9~ks, while during the last two observations 
51 counts were obtained in 257.9~ks. The Poisson probability of obtaining 
the last two observations (or any with a higher number of counts) 
given the first four is $\sim 5\times 10^{-11}$; the variability is
thus highly significant even considering
issues such as those described by Press \& Schechter (1974). 
We are not aware of any instrumental effects that might explain the
observed variability, and only $\approx 4.0$ background counts are
expected in the aperture (see \S2) during the entire 479.7~ks exposure. 
We note that the much brighter source CXOHDFN~J123646.3+621404
does not show significant evidence for variability; apparent variability of
this source might have been expected if the X-ray detection efficiency 
had changed substantially between observations.\footnote{See 
http://www.astro.psu.edu/users/townsley/cti/I3memo/
for constraints on temporal changes of the X-ray detection efficiency 
when radiation damage correction is performed with the 
procedure of Townsley et~al. (2000).}
The observed count-rate variability by a factor of $\approx 3$ occurred over 
$\approx 2.7$ months in the object's rest frame; CXOHDFN~J123639.5+621230
is one of the highest redshift X-ray variable objects known. The amplitude
of the observed X-ray variability appears consistent with that seen 
from low-redshift AGN of similar X-ray luminosity over comparable timescales 
(e.g., McHardy, Papadakis, \& Uttley 1998). 

{\it CXOHDFN~J123643.9+621249:}
This \chandra\ source is associated with a $z=0.557$ AGN candidate (see H01).
The K-S test suggests count-rate variability of this source at the 98.5\%
confidence level (see Table~2). Examination of the soft-band light curve in 
Figure~6 indicates a probable drop in count rate during the last two 
observations. During the first four observations, we obtained seven 
source plus background counts in 221.9~ks, while during the last two 
observations we obtained zero counts in 257.9~ks. The Poisson probability 
of obtaining the last two observations given the first four is 
$\sim 3\times 10^{-4}$, so the variability is considered to be significant 
at the $\approx 3.5\sigma$ level. The required amplitude of count-rate
variability is poorly constrained, but we estimate it to be at least
a factor of $\approx 2$. 
This apparent X-ray variability (combined with the observed luminosity)
supports the AGN nature of CXOHDFN~J123643.9+621249. The amplitude and 
timescale of the variability are consistent with what is seen from 
low-redshift AGN of similar luminosity (e.g., McHardy et~al. 1998). 

{\it CXOHDFN~J123646.3+621404:}
We have now collected sufficient counts from this $z=0.960$ broad-line 
AGN for moderate-quality spectral fitting. In this fitting, we have 
treated separately the data taken when the focal plane temperature 
was $-110^{\circ}$~C and $-120^{\circ}$~C; the spectra and response
matrices were created following G01. 
A simple power-law model with only 
Galactic absorption clearly fails to fit the spectrum
and is statistically rejected with $>99.9$\% confidence. The 
resulting residuals strongly suggest the presence of intrinsic
X-ray absorption (note also the fairly large band ratio of this
source; see Table~2); when neutral intrinsic absorption
is added we obtain a good fit with 
$\chi^2=31.9$ for 37 degrees of freedom (see Figure~7). The 
derived intrinsic column density is 
$(3.97^{+1.41}_{-0.94})\times 10^{22}$~cm$^{-2}$, and the photon 
index is $\Gamma=1.57^{+0.26}_{-0.22}$ (fit parameter errors are
quoted for $\Delta\chi^2=2.71$ here and hereafter); the photon 
index, while rather flat, is consistent with those seen from other 
AGN of comparable X-ray luminosity (e.g., Brandt, Mathur, \& Elvis 1997). 
The addition of ionized intrinsic absorption can also provide
a good fit to the data, although the fit parameters are
poorly constrained. 

The large intrinsic column density derived for the X-ray source has 
prompted us to search for signs of absorption at other wavelengths: 
e.g., occultation of the Broad Line Region by a torus or absorption
in the rest-frame ultraviolet (UV). Phillips et~al. (1997) have described 
a Keck spectrum of this object, and we have obtained and analyzed this 
spectrum via the DEEP project 
database.\footnote{See http://deep.ucolick.org/hdf/hdf.html}
Two narrow Mg~{\sc ii} absorption doublets are seen 
superposed upon the blue wing of the broad (FWHM~$\approx 6700$~km~s$^{-1}$) 
Mg~{\sc ii} emission line (see Figure~8 and Phillips et~al. 1997). 
For the stronger doublet, the rest-frame equivalent widths (EWs) of 
the Mg~{\sc ii}~$\lambda 2796.35$ and Mg~{\sc ii}~$\lambda 2803.53$ 
absorption lines are $2.7\pm 0.4$~\AA\ and $3.1\pm 0.4$~\AA, 
respectively. For the weaker doublet, the corresponding EWs are
$1.4\pm 0.3$~\AA\ and $1.2\pm 0.4$~\AA. While the breadth of the 
observed Mg~{\sc ii} emission line demonstrates that the 
Broad Line Region is not hidden by a torus (i.e., this is not a 
Type~2 AGN), there is clearly absorbing material along the line 
of sight. 

Intrinsic X-ray and UV absorption are often seen together in the spectra
of Seyfert galaxies and quasars, although the precise relation between
these two forms of absorption is still debated (e.g., Crenshaw et~al. 1999; 
BLW; and references therein). The large column density derived for the 
X-ray absorber strongly suggests that it is related to the AGN. The UV 
Mg~{\sc ii} absorption is also likely to be intrinsic given the large 
doublet EWs and the fact that there are two absorption doublets so
close together (compare with Steidel \& Sargent 1992 
and Aldcroft, Bechtold, \& Elvis 1994). The EW 
ratios of the Mg~{\sc ii} doublets are consistent with both of them being 
saturated, and it is not possible to derive reliable Mg~{\sc ii}
column densities from these data. The outflow velocity for the 
stronger Mg~{\sc ii} doublet is $\approx 230$~km~s$^{-1}$, and
the outflow velocity for the weaker doublet is $\approx 1700$~km~s$^{-1}$; 
these velocities are measured relative to the systemic redshift of
$z=0.960$ rather than the poorly defined peak of the Mg~{\sc ii}
emission line (see Appendix~B of Phillips et~al. 1997). It is possible that 
both the X-ray and UV absorption arise in an outflowing ``warm absorber'' 
similar to those seen in Seyfert galaxies at low redshift. Mg~{\sc ii}
absorption is not commonly seen from Seyfert galaxies with warm absorbers, 
but it is detected in some such as NGC~4151 (e.g., Crenshaw et~al. 1999).

Although this AGN is known to be variable in the radio
(Richards et~al. 1998; by $\approx 30$\% over an 18 month period) 
and optical
(Sarajedini et~al. 2000; by $\approx 8$\% over an 24 month period) 
and is by far the brightest X-ray source in the HDF-N, no significant 
X-ray variability is detected. 

{\it CXOHDFN~J123651.7+621221:} 
This source is coincident with the remarkably red object \nicmos\ 
(Dickinson et~al. 2000; also see \S4 of Dickinson 2000) which has 
been argued to be either a high-redshift starburst enshrouded 
in dust (e.g., Muxlow et~al. 1999) or a Type~2 quasar candidate (H00). 
With the increased exposure time, this source is now detected in the 
ultrahard band from 4--8~keV with $9.73\pm 3.32$ counts. If 
this source is indeed at moderate or high 
redshift (see Table~3), it almost certainly contains an active 
nucleus given its large hard X-ray luminosity and its hard X-ray
spectral shape. Note that it has the largest band ratio of any source
in the HDF-N (see Table~2).  

Although this object is the second brightest X-ray source in the HDF-N
and is of great scientific interest, only crude X-ray spectral fitting
is possible at present. The large band ratio suggests the probable
presence of X-ray absorption, so we have fitted the spectrum with a 
$\Gamma=2$ power-law model that is absorbed by neutral gas at
$z=2.7$ (see Table~3). The resulting fit is acceptable although
poorly constrained, and the best-fit column density is 
$(4.27^{+2.27}_{-1.72})\times 10^{23}$~cm$^{-2}$. The corresponding
absorption-corrected luminosity in the 0.5--8.0~keV band is
$\approx 2\times 10^{44}$~erg~s$^{-1}$. 


\subsection{Band Ratios and X-ray-to-Optical Flux Ratios}

In Figure~9 we show a plot of X-ray band ratio (see Table~2) versus 
soft-band count rate. This compares the HDF-N 
sources detected here to the sources detected with 
221.9~ks of \chandra\ exposure by H01 (also compare with Figure~2 of 
Giacconi et~al. 2001). Our faintest HDF-N sources occupy a region of 
low count rate and relatively low band ratio that has not been previously 
seen for individual sources, although stacking analyses had suggested
that sources would occupy this region (e.g., Figure~2 of 
Giacconi et~al. 2001). This is partially due to the fact that, with 
more exposure, we can place tighter upper limits on the X-ray band 
ratios for several of these sources. 

The two hardest X-ray sources in the HDF-N itself are 
CXOHDFN~J123646.3+621404 and CXOHDFN~J123651.7+621221;
notably, these are also the two brightest HDF-N sources in terms 
of their full-band emission (see Table~2). In \S3.2 it was shown
that the hard spectrum of CXOHDFN~J123646.3+621404 is almost 
certainly due to intrinsic X-ray absorption, and this is 
probably the case for CXOHDFN~J123651.7+621221 as well. 

In Figure~10 we plot $R$ magnitude versus soft-band flux. All but 
one of the faintest ($\simlt 2\times 10^{-16}$~erg~cm$^{-2}$~s$^{-1}$)
\chandra\ sources in the HDF-N lie below the region occupied by
luminous AGN, as expected from the source descriptions in \S3.1 
and \S3.2. Some of these sources appear to be low-luminosity AGN 
while others are normal galaxies. 
The one faint X-ray source that is also optically faint is
CXOHDFN~J123642.0+621331, which occupies the AGN region
as expected from Figure~3. 

Although the X-ray faint, optically bright sources
comprise a heterogeneous population, we have determined an
average band ratio for these sources to constrain their average
spectral shape. We have used the six sources lying below the
$\log (f_{\rm X}/f_{\rm R})=-1$ line in Figure~10 as well as 
CXOHDFN~J123649.5+621345. The average band ratio is $0.21\pm 0.08$, 
corresponding to a photon index of $\Gamma=2.1\pm 0.1$ for a
power-law model with the Galactic column density. This photon 
index is consistent with that of X-ray binaries and some 
low-luminosity AGN, although we are hesitant to provide
detailed interpretation due to the heterogeneous nature of the 
averaged sources. A similar photon index has recently been derived 
by Tozzi et~al. (2001). 


\subsection{Stacking Analysis Results}

\subsubsection{Optically Bright Galaxies}

Several of the optically bright galaxies in the HDF-N have now been 
found to be low-luminosity X-ray sources (see Figure~2). In order 
to extend these results, a stacking analysis was performed 
to attempt to detect the HDF-N galaxies from W96 with 
$V_{606}<22.5$ that are not individually detected in X-rays. 
Two $V_{606}<22.5$ galaxies lying near (but unrelated to) relatively 
bright \chandra\ sources were removed from consideration to minimize 
background and avoid any contamination (sources 3-400.0 and 3-659.0 
in W96). The X-ray data for all $V_{606}<22.5$ galaxies were
manually inspected to ensure that none of them is an X-ray source 
lying just below the detection limit; none of the sources contains 
a point source with $\simgt 5$ full-band counts, and most would 
not admit a point source with $\simgt 3.5$ full-band counts
(after subtraction of the expected background). The 
resulting galaxy sample has 17 objects with $V_{606}$ from 
19.9--22.5 including 11 spirals, three ellipticals and three 
irregulars (see Table~4 and Figure~2). None of these galaxies 
is known to have broad optical emission lines (Cohen et~al. 2000), and 
none of them has been identified as an AGN candidate in the searches of 
Jarvis \& MacAlpine (1998), Conti et~al. (1999),  Liu et~al. (1999), 
Sarajedini et~al. (2000) and Vanden~Berk et~al. (2000). Only two of 
them (sources 2-404.0 and 2-736.0 in W96) have been detected in the radio by 
Richards et~al. (1998) and Richards (2000). 
The effective exposure time of the stacked image is 7.7~Ms, 
and we have used the restricted ACIS grade set in this analysis.  

If the galaxies under study are indeed X-ray sources, some of them are 
likely to have their X-ray emission offset from their nuclei (as for the 
cases of CXOHDFN~J123641.7+621131 and CXOHDFN~J123649.5+621345), 
but without a priori knowledge of the location of this emission we are
limited to stacking these galaxies at their nuclei
(with one exception; see Table~4). Given 
the optical extents of these galaxies, X-ray emission 
up to $\approx 1.5^{\prime\prime}$ from the nucleus 
is plausible in most cases. Therefore, for each galaxy the 
30 pixels with centers within $1.5^{\prime\prime}$ of the 
stacking position were considered (see Figure~11a). 
In the full-band stacked image, 79 counts 
were detected in these 30 pixels
while 49.7 are expected from background. The Poisson probability of
obtaining 79 counts or more when 49.7 are expected is only 
$8\times 10^{-5}$, so full-band X-ray emission from the average 
galaxy is detected with high significance. Figure~11b shows
that the full-band X-ray emission appears to be well centered 
on the nuclear positions of the galaxies, although the centroid
position is in fact uncertain at the $\approx 1^{\prime\prime}$ 
level due to the combination of limited photon statistics, 
background, and the ACIS pixel size. 
There also appears to be a detection in the stacked soft-band image. 
Following the same procedure as for the full-band image, we obtain
29 events when 10.5 are expected; the corresponding Poisson probability
is $2\times 10^{-6}$. The soft-band emission centroid appears somewhat 
offset from the nuclear positions (see Figure~11c), although as above
we note that the centroid position is fairly uncertain. 
We have not found statistically significant evidence for a detection
in the hard-band stacked image. 

Source searching with {\sc wavdetect} confirms the detections from the 
manual Poisson calculations above, although the detection significances
are somewhat lower. The full-band and soft-band sources are detected
when {\sc wavdetect} is run with probability thresholds of 
$5\times 10^{-5}$ and $2\times 10^{-5}$, respectively. Given 
that 30 pixels are being searched in each case, the respective
probabilities of false detections are $\approx 1.5\times 10^{-3}$ 
and $\approx 6\times 10^{-4}$. 
Monte-Carlo simulations were also performed to examine the reliability 
of the stacking analyses; good agreement was obtained with the Poisson 
statistics calculations above (see Appendix~A).  
Finally, these stacking analyses are not subject to biases 
such as those discussed by Eddington (1913) because the number of sources 
under consideration is fixed and was not determined using the X-ray data. 

One concern with the interpretation of an average brightness from a stacking
analysis is that simply dividing the observed number of counts by the number 
of sources does not necessarily yield an accurate estimate of the characteristic
brightness of the population; e.g., the bulk of the observed counts might
be produced by only one or two sources. In this study, however, we believe
that averages do reasonably reflect typical brightnesses. In the 
full-band stacking above, for example, there are $\approx 29.3$ observed counts 
from 17 sources. No individual source, as noted above, has $\simgt 5$ counts. 
Therefore at least $\approx 6$ sources, and very likely more, are contributing to 
the observed counts.

We have repeated the stacking analysis above for only the 11 spirals
with $V_{606}<22.5$ (see Table~4); the effective exposure time of the 
resulting stacked image is 5.0~Ms. In the full-band image we obtain 47 
events when 32.1 are expected (see Figure~11d); the corresponding Poisson 
probability is $8\times 10^{-3}$, and the Monte-Carlo analysis in Appendix~A 
gives essentially the same probability. This detection is significant
at the 99.2\% confidence level. We do not obtain significant detections 
in the soft band or hard band for the spiral galaxy sample. 

\subsubsection{Radio Sources}

In H00 we described the fairly good correspondence between X-ray sources 
and radio sources, and this continues now that deeper X-ray data are 
available. For example, eight of our 12 \chandra\ sources in the HDF-N 
are detected at 8.5~GHz by Richards et~al. (1998). Furthermore,
we preferentially detect in X-rays the brighter 8.5~GHz sources: 
all but one of the six radio sources with 8.5~GHz fluxes above 10~$\mu$Jy
are detected, compared to only three of the nine radio sources with 8.5~GHz 
fluxes below 10~$\mu$Jy. 
To constrain the X-ray properties of the seven 8.5~GHz HDF-N 
sources not detected individually in X-rays, we have stacked them
but do not obtain statistically significant point-source detections in 
any of our bands. The upper limit on a point source in the full-band
stacked image is $\approx 6.8$ counts in 3.2~Ms (this is a 95\% confidence
upper limit calculated following Kraft, Burrows, \& Nousek 1991). Adopting
the spectral model used in \S2.2, the upper limit on the full-band flux
of the average source is $\approx 2.1\times 10^{-17}$~erg~cm$^{-2}$~s$^{-1}$. 


\section{Discussion}

\subsection{General Source Properties and Number Counts}

With the new data, the number of \chandra\ sources detected in the HDF-N 
itself has risen from eight (H01) to 12. Nine of these 12 sources are 
detected in the full band, 11 are detected in the soft band, six are 
detected in the hard band, and two are detected in the ultrahard band. 
We have not found any sources that are detected only in the hard or
ultrahard bands. 

All of the \chandra\ HDF-N sources have likely counterparts at 
optical or near-infrared wavelengths. Most of the \chandra\ sources
reside in optically luminous galaxies (see Table~3), and at a given
redshift they tend to be among the optically brightest galaxies in
the HDF-N (for our cosmology an $L^\ast$ galaxy has 
$M_{\rm B}\approx -20.5$). This result is consistent with
that of Barger et~al. (2001), extending it downward in X-ray
flux by a factor of $\approx 5$. There are, however, some apparent
exceptions to this trend. For example, if CXOHDFN~J123656.9+621301
is at $z=0.474$ (see Table~3), it is one of the optically fainter
galaxies at this redshift. We also note that the hosts of both
CXOHDFN~J123641.7+621131 and CXOHDFN~J123649.5+621345 have 
fairly low optical luminosities, and in each of these cases
the X-ray source appears to be spatially offset from the nucleus
(and thus is presumably not associated with a supermassive black
hole). It will be important to avoid such objects when using deep
X-ray surveys to conduct censuses of supermassive black holes
in the local ($z\simlt 0.5$) Universe. This can probably be done
with careful astrometric analysis and by exploiting the large 
negative $\log (f_{\rm X}/f_{\rm R})$ values of these 
objects (see \S3.3). 

Most of the \chandra\ HDF-N sources have radio luminosities consistent with
those of normal galaxies (e.g., Condon 1992); this is in agreement 
with the results of Barger et~al. (2001) and H01, and it extends these
results to fainter X-ray flux levels. The only \chandra\ sources that 
appear to be radio-loud AGN (using the criterion of Hooper et~al. 1996) 
are CXOHDFN~J123642.0+621331 (provided its redshift is $z=4.424$; see \S3.1) 
and perhaps the FR~I galaxy CXOHDFN~J123644.3+621132 (see Figure~4). 
CXOHDFN~J123646.3+621404 and CXOHDFN~J123651.7+621221 also have 
enhanced radio emission compared to that expected from a normal 
galaxy, although they are not radio loud. 
We have compared the radio angular sizes and morphologies of our \chandra\ 
sources to those of the other radio sources in 
Table~3 of Richards et~al. (1998), 
Chapter~4 of Richards (1999), and 
Table~2 of Richards (2000); one might expect X-ray emitting AGN
to have core-jet or compact morphologies.  
CXOHDFN~J123644.3+621132 has a clear core-jet radio morphology
(see Figure~4), and CXOHDFN~J123646.3+621404 and CXOHDFN~J123642.0+621331
are notably compact with 8.5~GHz angular sizes $<0.1^{\prime\prime}$. 
However, due to the limited sample size and radio constraints, it is
not possible to draw general conclusions at present. 

The corresponding soft-band and hard-band 
source densities for this 5.3~arcmin$^2$ field are 
$7500\pm 3000$~deg$^{-2}$ (at \hbox{$\approx 4.9\times 10^{-17}$~erg~cm$^{-2}$~s$^{-1}$}) and 
$4100\pm 2400$~deg$^{-2}$ (at \hbox{$\approx 2.3\times 10^{-16}$~erg~cm$^{-2}$~s$^{-1}$}), 
respectively (error bars on these values have been calculated following
Gehrels 1996). These flux limits are a factor of $\geq 2$ lower than those
previously reported (Mushotzky et~al. 2000; G01; 
Giacconi et~al. 2001; Hasinger et~al. 2001), 
and these source densities are the highest yet reported. The number counts 
in the soft band lie somewhat above an extrapolation of those cited above, 
although the discrepancy is not highly significant; any excess may be 
due to the emergence of the optically bright galaxies as low-luminosity 
X-ray sources (e.g., \S5.4 of H01). The number counts in the hard band are 
consistent with an extrapolation of those cited above. 

\subsection{Stacking Analyses of Optically Bright Galaxies}

A comparison of our HDF-N X-ray source list with the Cohen et~al. (2000)
and Cohen (2001) redshift catalogs for the HDF-N reveals that 
three of the five lowest-redshift ($z<0.15$) galaxies known in the 
HDF-N have now been detected in the X-ray band, 
including {\it all\/} of those with $M_{\rm V}<-18$ (the two non-detected 
galaxies at $z<0.15$ both have absolute magnitudes of only 
$M_{\rm V}\approx -15$). This high detection fraction is understandable
given that the sensitivity needed to detect the
most luminous ($\simgt 2\times 10^{39}$~erg~s$^{-1}$) X-ray binaries 
and supernova remnants at $z\simlt 0.15$ (e.g., Schlegel 1995; 
Makishima et~al. 2000) has now been achieved. Indeed, two 
of the three low-redshift X-ray detections 
are spatially offset from their host galaxies' 
nuclei. For comparison, our sensitivity limit would allow detection 
of the Galactic microquasar GRS~1915+105 (one of the most luminous 
X-ray sources in the Galaxy; e.g., Greiner, Morgan, \& Remillard 1996)
to $z\approx 0.1$. Similarly, we could detect the most luminous known 
off-nuclear source in M82 at its peak (e.g., Kaaret et~al. 2001) 
to $z\approx 0.5$.

The stacking analyses in \S3.4.1 have extended the constraints on 
galaxy X-ray emission to $z\approx$~0.2--1 for galaxies with 
$V_{606}=$~19.9--22.5. These analyses are complementary to the 
cross-correlation work performed with \rosat\ on galaxies with
similar optical magnitudes (e.g., Almaini et~al. 1997 and 
references therein; also see Refregier, Helfand, \& McMahon 1997
for the cross-correlation work on optically brighter galaxies). 
Because of our much greater sensitivity, a positive signal 
is obtained with many fewer galaxies (the \rosat\ 
cross-correlations needed $\sim 5000$ galaxies while we used 11--17). 
The greater sensitivity of this study also provides much tighter 
constraints on the X-ray luminosities of the individual galaxies used 
in the analysis. For comparison, the 0.5--2.0~keV luminosity
limit for an individual galaxy at $z=0.7$ is 
$\approx 9\times 10^{40}$~erg~s$^{-1}$ in the current study
while the \rosat\ cross-correlation studies had individual galaxy 
luminosity limits of $\approx 8\times 10^{42}$~erg~s$^{-1}$ at this 
redshift. Unlike the \rosat\ cross-correlation studies, we should be 
fairly immune to contamination by Seyfert nuclei (e.g., see \S4 of 
Almaini et~al. 1997); note that our good 2--8~keV sensitivity should
also allow us to detect moderately obscured AGN with low-level
soft X-ray emission. It is still possible that some low-luminosity 
AGN remain in the galaxy sample in Table~4; low-luminosity AGN are
seen in $\sim 40$\% of local galaxies and have X-ray luminosities 
ranging from $\simlt 10^{38}$~erg~s$^{-1}$ to $10^{41}$~erg~s$^{-1}$ 
with a median luminosity of $\sim 4\times 10^{38}$~erg~s$^{-1}$
(e.g., Ho et~al. 2001). None of the galaxies in Table~4 has been 
identified as an AGN in the intensive HDF-N follow-up studies 
(see \S3.4.1), although the current optical spectroscopy probably 
would not have detected the subtle spectral features often 
associated with low-luminosity AGNs. 

Stacking the full-band images of the 11 spirals in Table~4 results in a 
detection of $\approx 14.9$ net counts in an effective exposure time of 
5.0~Ms. At the median redshift of $z=0.474$, the rest-frame 
bandpass corresponding to the full-band detection is 0.74--11.79~keV. 
Provided the basic X-ray production mechanisms in these spirals are similar 
to those of spirals in the local Universe, the emission in this bandpass is 
probably dominated by X-ray binaries and has an effective power-law
photon index of $\Gamma\approx 2.0$ (e.g., Kim, Fabbiano, \& Trinchieri 1992). 
With this spectral model and the Galactic column density, 
the observed count rate implies a 0.5--8.0~keV flux of
$\approx~2.3\times 10^{-17}$~erg~cm$^{-2}$~s$^{-1}$ for the average 
spiral in our sample (the corresponding 0.5--2.0~keV flux is 
$\approx~1.1\times 10^{-17}$~erg~cm$^{-2}$~s$^{-1}$).\footnote{We have
corrected for the use of the restricted ACIS grade set rather than
the standard \asca\ grade set using the count ratios determined for
CXOHDFN~J123646.3+621404 in \S2.1.} At the median redshift of 
$z=0.474$, the implied rest-frame 0.5--8.0~keV luminosity is 
$\approx 1.6\times 10^{40}$~erg~s$^{-1}$. 

Normal spirals in the local Universe ($z<0.01$) with optical 
luminosities similar to those in the stacking sample have X-ray 
luminosities in the range $\approx 2\times 10^{39}$~erg~s$^{-1}$ to 
$\approx 2\times 10^{40}$~erg~s$^{-1}$. For example, the 
11 spirals in Fabbiano, Trinchieri, \& MacDonald (1984) with 
$M_{\rm B}<-18$ and X-ray detections have an average 0.5--8~keV 
luminosity of $6.4\times 10^{39}$~erg~s$^{-1}$.\footnote{We have
corrected the luminosities from Fabbiano et~al. (1984) for bandpass
effects and have converted them into our adopted cosmology.}
Our derived luminosity of $\approx 1.6\times 10^{40}$~erg~s$^{-1}$
for the average $z\approx 0.5$ spiral is somewhat higher than 
for the $z<0.01$ Fabbiano et~al. (1984) spirals. However, the
two samples are not inconsistent in the amount of X-ray
luminosity per unit $B$-band luminosity. The average spiral 
used in our stacking analysis is $\approx 2$ times more 
luminous in the $B$ band than the average spiral in the 
Fabbiano et~al. (1984) comparison sample (although the 
medians of the two $B$-band luminosity distributions only 
differ by a factor of $\approx 1.3$). 


At the median redshift of $z=0.474$, the lookback time is 
$\approx {1\over 3}$ of the Hubble time. Thus, if this sample 
is representative of the Universe as a whole, it appears that 
the X-ray luminosities of normal spirals (per unit $B$-band 
luminosity) have not evolved upward by more than a factor 
of $\approx 2$ over the past $\approx 4.5$~Gyr. 

White \& Ghosh (1998) considered the effects of the declining cosmic
star-formation rate on the evolution of low-mass X-ray binary 
(LMXB) populations in galaxies. They suggested that LMXB activity
should peak at $z\approx$~0.5--1 (rather than coincident with the
peak in the cosmic star-formation rate) because the 
evolutionary timescale for LMXBs is a non-negligible fraction of 
the Hubble time. They also estimated that the X-ray luminosities 
of $z\approx$~0.5--1 galaxies should be at least an order of 
magnitude higher than they are today. While our current data
allow the average galaxy X-ray luminosity at $z\approx 0.5$ 
to be up to a factor of $\approx 2$ times higher than at $z<0.01$, 
we do not see any evidence for an order-of-magnitude luminosity
increase at this redshift. These are the first significant
constraints on models such as those proposed by White \& Ghosh (1998). 

A.~Ptak et~al., in preparation, have recently refined the X-ray luminosities
predicted by the LMXB evolution models of White \& Ghosh (1998). They
have made specific predictions for the \hbox{HDF-N} using the redshift data 
of Cohen et~al. (2000). They predict smaller X-ray luminosities for 
$z\approx$~0.5--1 galaxies than discussed by White \& Ghosh (1998), 
and their predicted luminosities 
[$\approx$~(3--4)$\times 10^{40}$~erg~s$^{-1}$ in the 0.5--8.0~keV band]
are closer to those found by our stacking analysis. 

\subsection{Narrow Emission-Line Galaxies}

The new \chandra\ data also allow 
us to test some predictions made previously
about X-ray sources in the HDF-N. Almaini \& Fabian (1997), for
example, made predictions of the number of AGN in the HDF-N 
and proposed that X-ray luminous narrow-emission line galaxies (NELGs) 
may comprise up to $\approx 10$\% of the $\approx 3000$ galaxies in 
the HDF-N. At our current soft-band sensitivity level, they predict
$\approx 9$ NELGs to be present. Six of the 12 HDF-N sources in Table~3 
have some narrow emission lines in their optical spectra (we exclude 
broad-line AGN and objects with absorption-dominated spectra). Two 
of these (CXOHDFN~J123641.7+621131 and 
CXOHDFN~J123648.3+621426), however, have X-ray luminosities below 
those typically associated with the NELG population (their X-ray
to optical flux ratios are more representative of those of
normal galaxies). Thus, we presently detect at most four X-ray 
luminous NELGs in the HDF-N (CXOHDFN~J123643.9+621249,  
CXOHDFN~J123648.0+621309, CXOHDFN~J123649.7+621312 and 
CXOHDFN~J123656.9+621301), below the predictions 
of Almaini \& Fabian (1997) and inconsistent with them 
at the $\approx 95$\% confidence level. Higher quality 
optical spectroscopy of these four objects may well reveal 
that some of them contain previously unrecognized AGN. 


\section{Summary}

We have used a 479.7~ks \chandra\ image, the deepest X-ray observation ever 
presented, to study the X-ray source content of the HDF-N and its immediate 
vicinity. The key results are the following: 

\begin{itemize}

\item
Four new X-ray sources in the HDF-N itself have been discovered bringing
the total number of HDF-N X-ray sources to 12. Two of the new sources
are optically bright galaxies at low redshift; with their detection
X-ray emission has now been detected from all three of the optically luminous 
($M_{\rm V}<-18$) HDF-N galaxies with $z<0.15$. The other two new X-ray
sources are an FR~I radio galaxy and an edge-on spiral that 
hosts a powerful starburst and/or low-luminosity AGN. 
See \S3.1 and \S4.1. 

\item
The remarkable $\mu$Jy radio source \vla\ is detected by \chandra. This 
source has a likely redshift of $z=4.424$, and this X-ray detection 
supports the presence of an AGN in this object. The \aox\ value of 
\vla\ is in good agreement with \aox\ values found for low-redshift
AGN. See \S3.1. 

\item
X-ray variability is found from the $z=3.479$ broad-line AGN CXOHDFN~J123639.5+621230 
and the $z=0.557$ AGN candidate CXOHDFN~J123643.9+621249. The first brightens
by a factor of $\approx 3$ over $\approx 2.7$ months in the rest frame, 
while the second dims by a factor of $\simgt 2$ in a rest-frame time interval
of $\approx 8$ months. The X-ray variability of CXOHDFN~J123643.9+621249
strongly supports the presence of an AGN. See \S3.2. 

\item
We have modeled the X-ray spectrum of the brightest X-ray source in the
HDF-N, the \hbox{$z=0.960$} broad-line 
AGN CXOHDFN~J123646.3+621404. Evidence is found 
for intrinsic X-ray absorption with a column density of 
$\approx 4\times 10^{22}$~cm$^{-2}$. This absorption is plausibly 
related to the Mg~{\sc ii} absorption doublets that are superposed 
upon the blue wing of the broad Mg~{\sc ii} emission line. See \S3.2. 

\item
Stacking analyses of $V_{606}<22.5$ galaxies not individually detected
in X-rays have provided estimates of the average X-ray fluxes of
these objects. Their X-ray emission probably arises from combinations
of X-ray binaries, supernova remnants, and low-luminosity AGN.
The average X-ray luminosity of normal spiral galaxies at $z\approx 0.5$ 
appears similar to that of spirals in 
the local Universe; this constrains models for the
evolution of LMXB populations in galaxies in response to the
declining cosmic star-formation rate. Monte-Carlo simulations demonstrate 
the validity of the stacking analyses and show that the \chandra\ ACIS
performs well at source detection even with effective exposure 
times of $\approx 8$~Ms. See \S3.4, \S4.2 and Appendix~A. 

\item
We have constrained the size of the population of X-ray luminous 
NELGs at the faintest X-ray fluxes. These sources are less common
than has been predicted. See \S4.3. 

\end{itemize}


\acknowledgments

This work would not have been possible without the enormous efforts 
of the entire \chandra\ and ACIS teams. 
We thank A.J.~Barger, F.E.~Bauer, J.C.~Charlton, C.W.~Churchill,
L.L.~Cowie and A.~Ptak for helpful discussions. 
We thank an anonymous referee for useful comments. 
We gratefully acknowledge the financial support of
NASA grant NAS~8-38252 (GPG, PI),
NSF CAREER award AST-9983783 (WNB, DMA),  
NASA GSRP grant NGT5-50247 and 
the Pennsylvania Space Grant Consortium (AEH), and
NSF grant AST99-00703~(DPS). 

%


\newpage
\appendix

\section{Monte-Carlo Testing of the Stacking Analyses} 

The stacking methods employed in \S3.4.1 have allowed us to study the
properties of full-band sources down to flux levels of 
$\approx 2.3\times 10^{-17}$~erg~cm$^{-2}$~s$^{-1}$, pushing the source
detection ability of \chandra\ beyond what has been achieved previously. 
As such, one must be cautious of instrumental effects and other potential 
problems that could lead to false source detections in the stacked images. 
In order to perform an ``end-to-end'' test of the stacking analysis
and empirically assess false-detection probabilities, we have performed
Monte-Carlo simulations. These were designed to reproduce the 
actual stacking performed as closely as possible. 

Consider the full-band stacking analysis of the 17 galaxies in 
Table~4 performed in \S3.4.1. To test this analysis, we have 
randomly selected 17 positions chosen to be at least 8 pixels
from known X-ray sources and to have similar background levels to 
those of the actual galaxies being stacked. We have then constructed 
a stacked image from the data around these 17 positions and
determined the number of counts in the 30 pixels with centers
within $1.5^{\prime\prime}$ of the stacking position (just as
was done at the actual galaxy positions; see \S3.4.1). Repeating 
the above 100,000 times, we have determined a distribution 
function showing the number of random trials giving a particular
number of counts. The result is shown in Figure~12a. The resulting
distribution is closely Gaussian; a Gaussian fit gives a mean of 
49.1 counts and a standard deviation of $\sigma=7.0$ counts. At the  
actual galaxy positions we obtained 79 counts. Only 9 of the 100,000
random trials yielded 79 counts or more, so the false-detection
probability is only $\approx 9\times 10^{-5}$. For comparison,
the Poisson statistics calculation in \S3.4.1 gave a false
detection probability of $8\times 10^{-5}$. The excellent agreement
between the Monte-Carlo and Poisson statistics calculations gives
us further confidence in the reliability of our stacking analyses
and argues that there are no subtle instrumental effects causing
false source detections. For example, subtle cosmic ray afterglows
in ACIS could not cause a false detection since they will not 
be preferentially aligned with optically bright HDF-N galaxies. 
The ACIS performs well at source detection even with effective 
exposure times of $\approx 8$~Ms. 

We have also performed Monte-Carlo testing of the other analyses
presented in \S3.4.1. The results are shown in Figures~12b and 12c, 
and we consistently obtain good agreement with the results presented
in \S3.4.1.



\newpage

\begin{deluxetable}{lllc}
\tabletypesize{\scriptsize}
\tablewidth{0pt}
\tablecaption {Grade Sets and Corresponding Background Levels \label{gradetable}}
\tablehead{
\colhead{Name}                            &                                
\colhead{Band}                            &                                
\colhead{Grades}                          &                                
\colhead{Background counts pixel$^{-1}$}                                            
}
\startdata
Standard \asca\ grade set  & Full      & \asca\ grades 0, 2, 3, 4, 6   & $1.1\times 10^{-1}$ \\
                           & Soft      & \asca\ grades 0, 2, 3, 4, 6   & $3.3\times 10^{-2}$ \\
                           & Hard      & \asca\ grades 0, 2, 3, 4, 6   & $7.8\times 10^{-2}$ \\
                           & Ultrahard & \asca\ grades 0, 2, 3, 4, 6   & $5.4\times 10^{-2}$ \\
                           &           &                               & \\
Restricted ACIS grade set  & Full      & ACIS grades 0, 2, 8, 16, 64   & $7.7\times 10^{-2}$ \\
                           & Soft      & ACIS grades 0, 64             & $2.1\times 10^{-2}$ \\
                           & Hard      & ACIS grades 0, 2, 8, 16       & $5.6\times 10^{-2}$ \\
                           & Ultrahard & ACIS grades 0, 2, 8, 16       & $3.6\times 10^{-2}$ \\
\enddata
%
%
\end{deluxetable}

\clearpage


\begin{deluxetable}{lllccccrrrc}
\rotate
\tabletypesize{\scriptsize}
\tablewidth{0pt}
\tablecaption {\chandra\ Properties of HDF-N Sources \label{chandraprops}}
\tablehead{
\multicolumn{2}{c}{Coordinates} &                                
\colhead{Det.}                  &
\multicolumn{3}{c}{Counts}      &
\colhead{Band}                  &
\multicolumn{3}{c}{Flux}        &
\colhead{Probability}           \\
\colhead{$\alpha_{2000}$}        &                                
\colhead{$\delta_{2000}$}        &                                
\colhead{Bands$^{\rm a}$}        &
\colhead{FB$^{\rm b}$}           &
\colhead{SB$^{\rm b}$}           &
\colhead{HB$^{\rm b}$}           &
\colhead{Ratio$^{\rm c}$}        &
\colhead{FB$^{\rm d}$}           &
\colhead{SB$^{\rm d}$}           &
\colhead{HB$^{\rm d}$}           &
\colhead{of Constancy$^{\rm e}$}       
}
\startdata
12 36 39.56 & +62 12 30.2           & FSH  & $65.96\pm 8.48$   & $57.10\pm 7.68$   & $11.59\pm 3.60$   & $0.20^{+0.07}_{-0.07}$ & 1.51    & 0.57    & 0.54     & $8.4\times 10^{-4}$ \\
12 36 41.77 & +62 11 31.8           & FS   & $19.94\pm 5.00$   & $14.86\pm 4.00$   & $<8.54$           & $<0.58$                & 0.45    & 0.14    & $<0.39$  & $3.3\times 10^{-2}$ \\
12 36 43.99 & +62 12 49.8           & S    & $<10.99$          & $5.43\pm 2.44$    & $<3.94$           & $<0.73$                & $<0.24$ & 0.05    & $<0.18$  & $1.5\times 10^{-2}$ \\
12 36 44.35 & +62 11 32.9           & S    & $<15.11$          & $9.28\pm 3.16$    & $<3.94$           & $<0.43$                & $<0.34$ & 0.09    & $<0.18$  & 0.76 \\
12 36 46.34 & +62 14 04.6           & FSHU & $646.97\pm 25.63$ & $302.56\pm 17.46$ & $347.77\pm 18.81$ & $1.15^{+0.09}_{-0.09}$ & 21.61   & 2.92    & 18.81    & 0.43 \\
12 36 48.04 & +62 13 09.1           & FSH  & $42.57\pm 7.00$   & $24.96\pm 5.09$   & $10.79\pm 3.60$   & $0.43^{+0.18}_{-0.16}$ & 0.82    & 0.24    & 0.47     & 0.45 \\
12 36 48.36 & +62 14 26.0           & S    & $<12.44$          & $6.35\pm 2.64$    & $<5.08$           & $<0.80$                & $<0.28$ & 0.06    & $<0.23$  & 0.56 \\
12 36 49.52 & +62 13 45.6           & F    & $4.55\pm 2.44$    & $<6.78$           & $<7.24$           & $\cdots$               & 0.11    & $<0.07$ & $<0.34$  & 0.94 \\
12 36 49.73 & +62 13 12.8           & FS   & $10.30\pm 3.60$   & $9.22\pm 3.16$    & $<5.08$           & $<0.55$                & 0.23    & 0.09    & $<0.23$  & 0.17 \\
12 36 51.73 & +62 12 21.3           & FSHU & $77.41\pm 9.05$   & $30.81\pm 5.65$   & $47.18\pm 7.07$   & $1.53^{+0.41}_{-0.33}$ & 2.97    & 0.29    & 2.68     & 0.42 \\
12 36 55.46 & +62 13 11.0           & FSH  & $60.47\pm 8.06$   & $44.74\pm 6.78$   & $17.92\pm 4.47$   & $0.40^{+0.12}_{-0.11}$ & 1.13    & 0.44    & 0.76     & 0.63 \\
12 36 56.92 & +62 13 01.5           & FSH  & $52.72\pm 7.54$   & $41.03\pm 6.48$   & $9.88\pm 3.31$    & $0.24^{+0.09}_{-0.09}$ & 0.84    & 0.41    & 0.40     & 0.11 \\
            &                       &      &                   &                   &                   &                        &         &         &          & \\
12 36 42.09 & +62 13 31.2$^{\rm f}$ & FS   & $10.60\pm 3.60$   & $8.33\pm 3.00$    & $<6.36$           & $<0.77$                & 0.24    & 0.08    & $<0.29$  & 0.11 \\
\enddata
\tablenotetext{a}{Bands in which this source is detected. ``F'' indicates 
full band, ``S'' indicates soft band, ``H'' indicates hard band, and
``U'' indicates ultrahard band. With two exceptions, detections 
are at the $1\times 10^{-7}$ false-positive probability level using either 
the standard \asca\ grade set or the restricted ACIS grade set. The first exception is the full-band detection 
of CXOHDFN~J123649.7+621312 which is at the $1\times 10^{-5}$ probability level for both the standard \asca\ 
grade set and the restricted ACIS grade set. The second exception is the full-band detection of 
CXOHDFN~J123642.0+621331 which is at the $1\times 10^{-5}$ probability level for the restricted ACIS 
grade set. Because both of these exceptions are detected at the $1\times 10^{-7}$ probability level in
the soft band, we are confident of their reality (see \S2).}
\tablenotetext{b}{Source counts and errors are as computed by {\sc wavdetect}, and these have been 
verified using manual photometry. Upper limits are calculated using the Bayesian method of 
Kraft, Burrows, \& Nousek (1991) for 95\% confidence; the uniform prior used by these authors 
results in fairly conservative upper limits, and other reasonable choices of priors do not 
materially change our scientific results. All values are for the standard \asca\ grade set, 
and we have not corrected these values for vignetting. ``FB'' indicates full band, 
``SB'' indicates soft band, and ``HB'' indicates hard band.}
\tablenotetext{c}{Defined as the ratio of counts between the 2--8~keV and 0.5--2.0~keV bands. Errors 
for this quantity are calculated following the ``numerical method'' described in \S1.7.3 of Lyons (1991).
These values have been corrected for differential vignetting between the hard band and soft band.}
\tablenotetext{d}{Fluxes are in units of $10^{-15}$~erg~cm$^{-2}$~s$^{-1}$. They have
been corrected for vignetting but are not corrected for Galactic absorption. They have 
been computed using the method detailed in \S3.3 of H01.
%
%
After consideration of instrumental systematics, spectral shape uncertainties, and photon 
statistics, we estimate flux errors to range from $\approx 15$\% for our brightest 
sources to $\approx 60$\% for our faintest sources (see \S3.3 of H01). 
``FB'' indicates full band, ``SB'' indicates soft band, and ``HB'' 
indicates hard band.}
\tablenotetext{e}{The probability, as derived from a Kolmogorov-Smirnov test, that the source count rate
is consistent with a constant value (see \S2.3). Testing has been done in the band where a 
source has its highest signal-to-noise ratio.}
\tablenotetext{f}{This source is located just outside the HDF-N.}
\end{deluxetable}


\begin{deluxetable}{llllcccrrrccl}
\rotate
\tabletypesize{\scriptsize}
\tablewidth{0pt}
\tablecaption {Identifications of HDF-N \chandra\ Sources \label{identifications}}
\tablehead{     
\multicolumn{2}{c}{Coordinates}        &                                       
\colhead{W96}                          &
\colhead{CXO/W96}                      &
\colhead{}                             &
\colhead{}                             &
\colhead{}                             &
\multicolumn{3}{c}{$\log(L_{\rm X})$}  &
\colhead{Other}                        &
\colhead{Spect.}                       &
\colhead{}                             \\
\colhead{$\alpha_{2000}$}                     &                                
\colhead{$\delta_{2000}$}                     &                                
\colhead{Name}                                &
\colhead{Offset($^{\prime\prime}$)$^{\rm a}$} &
\colhead{$R^{\rm b}$}                         &
\colhead{Redshift$^{\rm c}$}                  &
\colhead{$M_{\rm B}^{\rm d}$}                 &
\colhead{FB$^{\rm e}$}                        &
\colhead{SB$^{\rm e}$}                        &
\colhead{HB$^{\rm e}$}                        &
\colhead{$\lambda^{\rm f}$}                   &
\colhead{Type$^{\rm g}$}                      &
\colhead{Notes}                  
}
\startdata
12 36 39.56 & +62 12 30.2           & 4-852.12   & 0.16            & 24.3    & 3.479  & $-21.9$       & 43.8    & 43.1    & 43.4     &      & $\cal{EQ}$ & Broad-line AGN \\ 
12 36 41.77 & +62 11 31.8           & 4-976      & 0.89$^{\rm h}$  & 19.9    & 0.089  & $-17.2$       & 39.9    & 39.4    & $<39.8$  & I    & $\cal{EI}$ & Face-on spiral \\
12 36 43.99 & +62 12 49.8           & 4-402.31   & 0.33            & 21.3    & 0.557  & $-20.6$       & $<41.3$ & 40.6    & $<41.2$  & R, I & $\cal{EI}$ & AGN candidate \\ 
12 36 44.35 & +62 11 32.9           & 4-752.1    & 0.29            & 22.7    & 1.050  & $-22.7$       & $<42.0$ & 41.4    & $<41.7$  & R    & $\cal{A}$  & FR~I \\
12 36 46.34 & +62 14 04.6           & 2-251      & 0.03            & 22.1    & 0.960  & $-20.9$       & 43.5    & 42.6    & 43.5     & R, I & $\cal{Q}$  & Broad-line AGN \\ 
12 36 48.04 & +62 13 09.1           & 2-121      & 0.31            & 21.0    & 0.476  & $-20.4$       & 41.7    & 41.2    & 41.5     & I    & $\cal{I}$  & Elliptical \\
12 36 48.36 & +62 14 26.0           & 2-537.111  & 0.49            & 18.8    & 0.139  & $-19.0$       & $<40.1$ & 39.4    & $<40.0$  & R, I & $\cal{E}$  & Elliptical \\
12 36 49.52 & +62 13 45.6           & 2-456.1111 & 1.33$^{\rm i}$  & 18.3    & 0.089  & $-18.4$       & 39.3    & $<39.1$ & $<39.8$  & I    & $\cal{A}$  & S0 or Elliptical \\ 
12 36 49.73 & +62 13 12.8           & 2-264.1    & 0.31            & 21.9    & 0.475  & $-19.5$       & 41.1    & 40.7    & $<41.1$  & R, I & $\cal{I}$  & Edge-on spiral \\ 
12 36 51.73 & +62 12 21.3           & $\cdots$   & 0.22            & 26.5    & 2.7    & $-20.9$       & 43.3    & 42.1    & 43.3     & R, I & $\cdots$   & NICMOS \\
12 36 55.46 & +62 13 11.0           & 3-180      & 0.10            & 23.3    & 0.968  & $-21.6$       & 42.5    & 42.1    & 42.4     & R, I & $\cal{A}$  & Elliptical \\
12 36 56.92 & +62 13 01.5           & 3-355      & 0.02            & 23.4    & 0.474  & $-18.0$       & 41.8    & 41.4    & 41.5     & R, I & $\cal{E}$  & Elliptical \\
 & & & & & \\
12 36 42.09 & +62 13 31.2$^{\rm j}$ & $\cdots$   & 0.15            & $>26.0$ & 4.424  & $-23.0$       & 43.2    & 42.4    & $<43.4$  & R, I & $\cal{E}$  & AGN/Star-forming \\ 
\enddata
\tablenotetext{a}{Offset between the \chandra\ source and the proposed W96 identification. 
For CXOHDFN~J123651.7+621221 we quote the offset from \nicmos, and for
CXOHDFN~J123642.0+621331 we quote the offset from \vla.}
\tablenotetext{b}{These are Kron-Cousins $R$ magnitudes (see Bessel 1990) calculated following 
\S4.2 of H01.} 
\tablenotetext{c}{Redshifts are from Cohen et~al. (2000) and Cohen (2001) unless noted otherwise.
The redshift of CXOHDFN~J123651.7+621221 is photometric and subject to significant 
uncertainty (see \S4 of Dickinson 2000; M.~Dickinson 2000, private communication;
D.M. Alexander et~al., in preparation).
The redshift of CXOHDFN~123656.9+621301 has been disputed by Fern\'andez-Soto et~al. (2001) 
who propose $z=1.27$.
The redshift of CXOHDFN~123642.0+621331 is discussed in \S3.1 and references therein.} 
\tablenotetext{d}{Absolute $B$ magnitudes have been calculated using $K$ corrections derived from
(1) the spectra of Coleman, Wu, \& Weedman (1980), and 
(2) a power-law model with an optical slope of $\alpha_{\rm o}=-0.5$ and typical AGN emission lines. 
We have used the infrared flux densities of Waddington et~al. (1999) and M.~Dickinson 2000, private 
communication to calculate the absolute $B$ magnitudes for CXOHDFN~J123642.0+621331 and 
CXOHDFN~J123651.7+621221, respectively.}
\tablenotetext{e}{Luminosities are given in the rest frame. ``FB'' indicates 
full band, ``SB'' indicates soft band, and ``HB'' indicates hard band.}
\tablenotetext{f}{Other wavelength bands in which these sources are detected.
R: Radio (Richards et~al. 1998; Richards 2000); 
I: Mid-infrared (Aussel et~al. 1999; D.M. Alexander et~al., in preparation; and references therein).}
\tablenotetext{g}{Optical spectral type from \S2.6 of Cohen et~al. (2000). 
``$\cal{E}$'' galaxies have spectra dominated by emission lines, 
``$\cal{A}$'' galaxies have spectra dominated by absorption lines, 
``$\cal{I}$'' galaxies have spectra of intermediate type between ``$\cal{E}$'' and ``$\cal{A}$'', and
``$\cal{Q}$'' galaxies have broad emission lines. 
See Cohen et~al. (2000) for details and explanations of the other spectral types.}
\tablenotetext{h}{The offset between the X-ray source and the spatially coincident off-nuclear 
``bright spot'' is $0.14^{\prime\prime}$. The bright spot is most clearly seen at short optical
wavelengths, and it may be an off-nuclear starburst region (H00) or background AGN.}
\tablenotetext{i}{The X-ray source is coincident with one of the optically brightest galaxies 
in the HDF-N, but it is offset from the galaxy's nucleus suggesting emission from an X-ray
binary (see \S3.1).}
\tablenotetext{j}{This source is located just outside the HDF-N.}
\end{deluxetable}


\begin{deluxetable}{lllccclc}
\tabletypesize{\scriptsize}
\tablewidth{0pt}
\tablecaption {Optically Bright HDF-N Galaxies Used in the Stacking Analyses \label{stacktable}}
\tablehead{
\colhead{Name$^{\rm a}$}          &                                
\colhead{$\alpha_{2000}^{\rm a}$} &                                
\colhead{$\delta_{2000}^{\rm a}$} &                                
\colhead{$V_{606}^{\rm a}$}       &                                
\colhead{$z^{\rm b}$}             &                                                      
\colhead{$M_{\rm B}^{\rm c}$}     &                                                      
\colhead{Morph. Type$^{\rm d}$}   & 
\colhead{Spect. Type$^{\rm e}$}                                                        
}
\startdata
1-34.0             &  12 36 44.57 & 62 13 04.6 &  22.11 & 0.485 & $-19.9$ & Spiral     &  $\cal{I}$    \\
1-87.0             &  12 36 45.85 & 62 13 25.8 &  21.49 & 0.321 & $-19.2$ & Spiral     &  $\cal{I}$    \\
2-264.0            &  12 36 49.37 & 62 13 11.2 &  21.67 & 0.477 & $-20.5$ & Elliptical &  $\cal{E}$    \\
2-404.0            &  12 36 51.08 & 62 13 20.7 &  19.92 & 0.199 & $-19.5$ & Spiral     &  $\cal{I}$    \\
2-652.0            &  12 36 51.78 & 62 13 53.7 &  21.94 & 0.557 & $-20.6$ & Spiral     &  $\cal{IE}$   \\
2-736.0            &  12 36 52.78 & 62 13 54.4 &  22.26 & 1.355 & $-22.6$ & Irregular  &  $\cal{EA}$   \\
3-350.0            &  12 36 53.90 & 62 12 54.0 &  21.88 & 0.642 & $-21.2$ & Spiral     &  $\cal{I}$    \\
3-386.0            &  12 36 50.25 & 62 12 39.7 &  21.20 & 0.474 & $-20.7$ & Spiral     &  $\cal{I}$    \\
3-534.0            &  12 36 58.76 & 62 12 52.3 &  21.58 & 0.321 & $-19.1$ & Spiral     &  $\cal{I}$    \\
3-610.1            &  12 36 56.65 & 62 12 45.6 &  21.30 & 0.518 & $-20.9$ & Spiral     &  $\cal{A}$    \\
3-965.0            &  12 36 57.48 & 62 12 10.5 &  22.23 & 0.665 & $-21.5$ & Elliptical &  $\cal{A}$    \\
4-241.1            &  12 36 47.04 & 62 12 36.8 &  21.40 & 0.321 & $-19.3$ & Irregular  &  $\cal{E}$    \\
4-402.0            &  12 36 44.19 & 62 12 47.9 &  21.13 & 0.555 & $-21.1$ & Irregular  &  $\cal{E}$    \\
4-550.0$^{\rm f}$  &  12 36 46.20 & 62 11 41.2 &  22.18 & 1.013 & $-22.9$ & Spiral     &  $\cal{E}$    \\ 
4-656.0            &  12 36 42.91 & 62 12 16.2 &  21.27 & 0.454 & $-20.5$ & Spiral     &  $\cal{I}$    \\
4-744.0            &  12 36 43.80 & 62 11 42.8 &  22.40 & 0.765 & $-22.1$ & Elliptical &  $\cal{A}$    \\
4-795.0            &  12 36 41.94 & 62 12 05.4 &  21.52 & 0.432 & $-20.1$ & Spiral     &  $\cal{I}$    \\
\enddata
\tablenotetext{a}{From W96.}
\tablenotetext{b}{Redshifts are from Cohen et~al. (2000) and Cohen (2001).} 
\tablenotetext{c}{Absolute $B$ magnitudes have been calculated using $K$ corrections derived from
the spectra of Coleman et~al. (1980).}
\tablenotetext{d}{Morphological classifications are from 
van~den~Bergh et~al. (1996), Franceschini et~al. (1998),
Abraham et~al. (1999a), Abraham et~al. (1999b), and
van~den~Bergh et~al. (2000).} 
\tablenotetext{e}{Optical spectral type from \S2.6 of Cohen et~al. (2000). 
``$\cal{E}$'' galaxies have spectra dominated by emission lines, 
``$\cal{A}$'' galaxies have spectra dominated by absorption lines, and
``$\cal{I}$'' galaxies have spectra of intermediate type between ``$\cal{E}$'' and ``$\cal{A}$''.
None of these galaxies is known to have broad optical emission lines.
See Cohen et~al. (2000) for details and explanations of the other spectral types.}
\tablenotetext{f}{This face-on spiral has a set of bright spots along its arms that
are brighter than its nucleus in the rest-frame UV. These are presumably star-forming 
regions, and we have centered the relevant \chandra\ image on the centroid of these.}
\end{deluxetable}

\clearpage


\begin{figure}
\vspace{-0.5truein}
\epsscale{0.9}
\figurenum{1}
\plotone{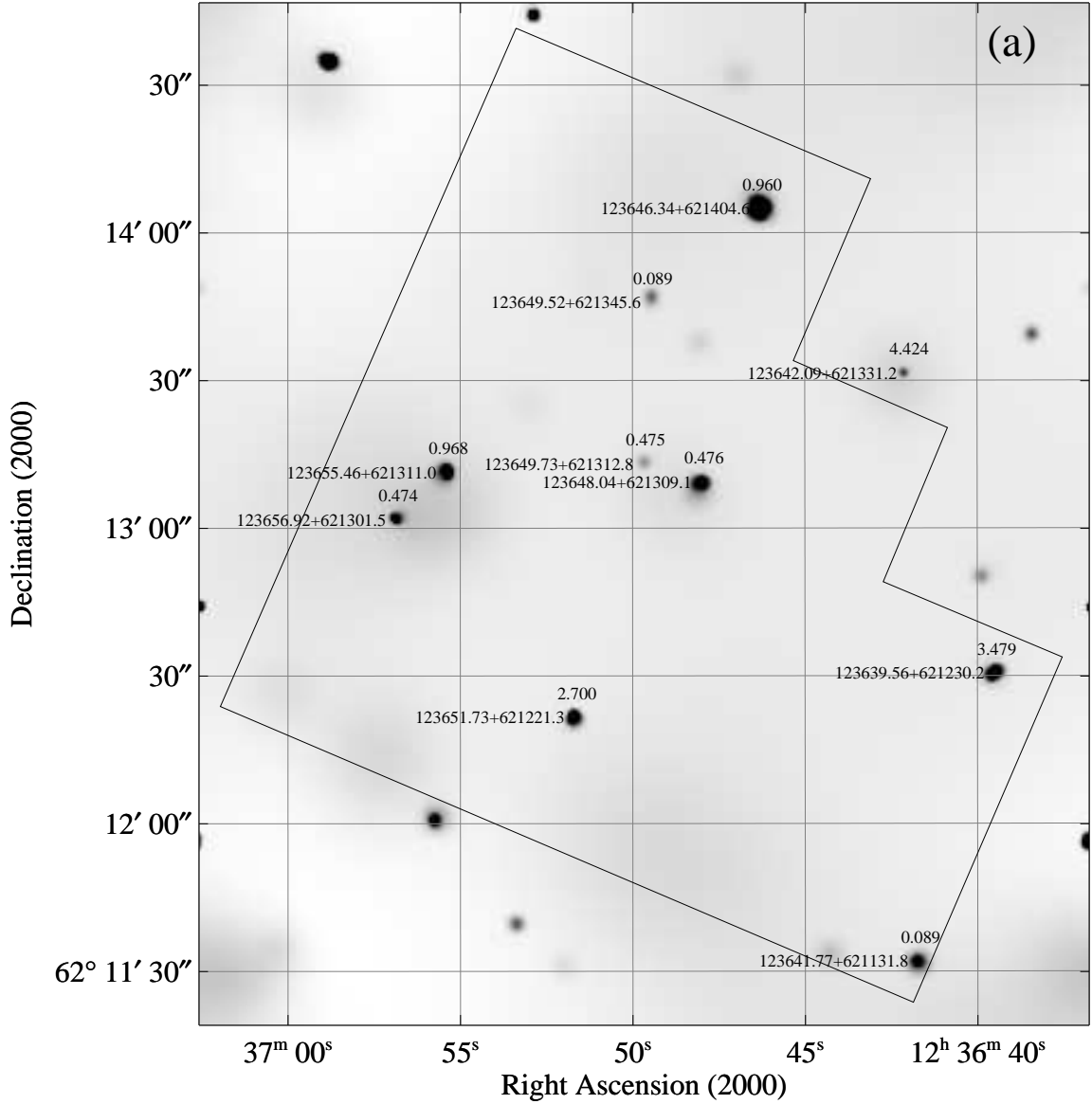}

\caption{Adaptively smoothed \chandra\ images of the HDF-N and its immediate
environs in the (a) full band, (b) soft band, and (c) hard band. Sources 
discussed in this paper are labeled; labels are given immediately to the left 
of the corresponding X-ray sources. Redshifts for these sources are 
given immediately above the corresponding X-ray sources. These 
images have been made using the restricted ACIS grade set, and the 
adaptive smoothing has been performed using the code of 
Ebeling, White, \& Rangarajan (2001) at the $2.5\sigma$ level. The 
grayscales are linear. 
\label{asmoothed}}
\end{figure}

\begin{figure}
\epsscale{0.9}
\figurenum{1b}
\plotone{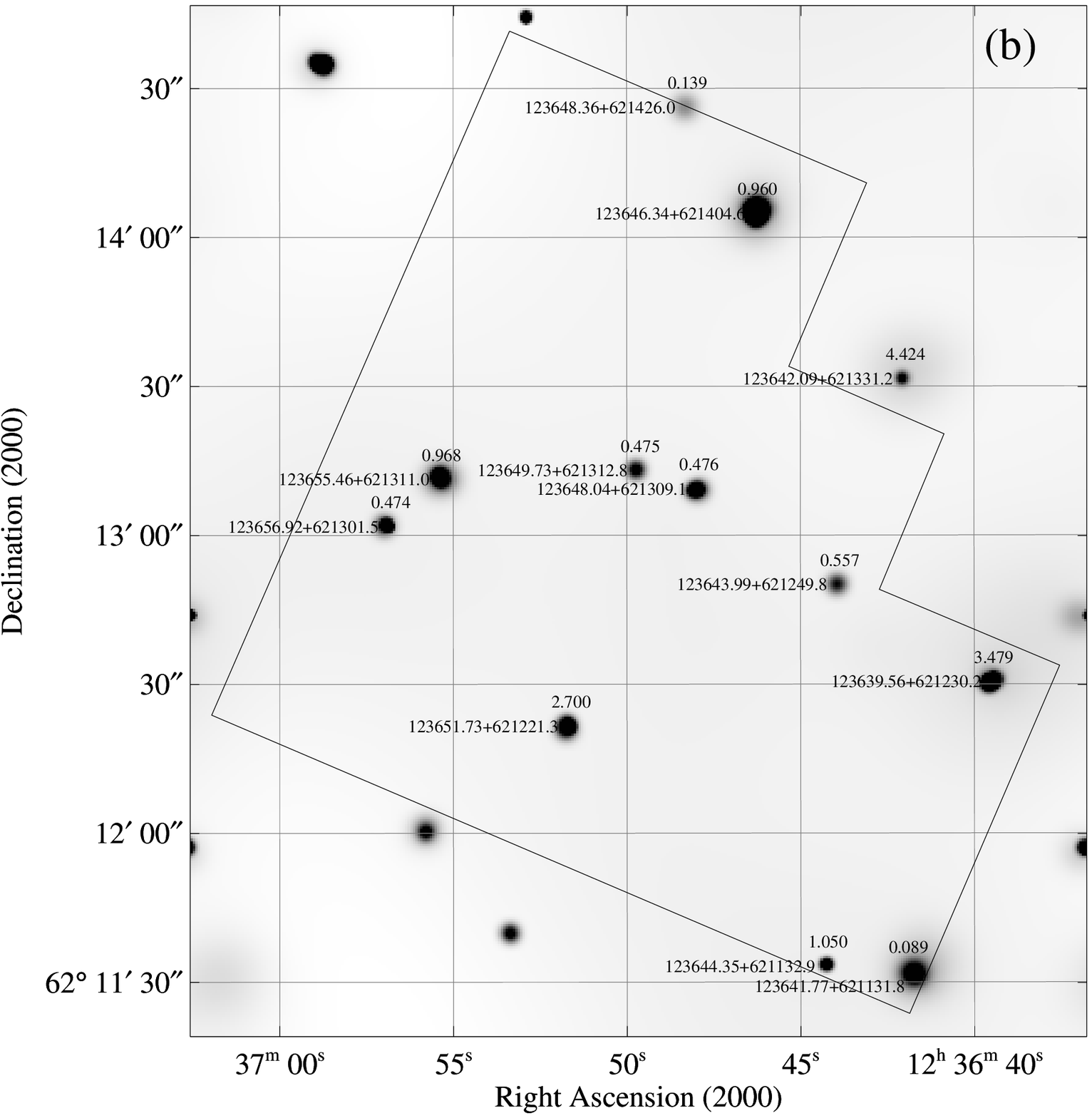}
\end{figure}

\begin{figure}
\epsscale{0.9}
\figurenum{1b}
\plotone{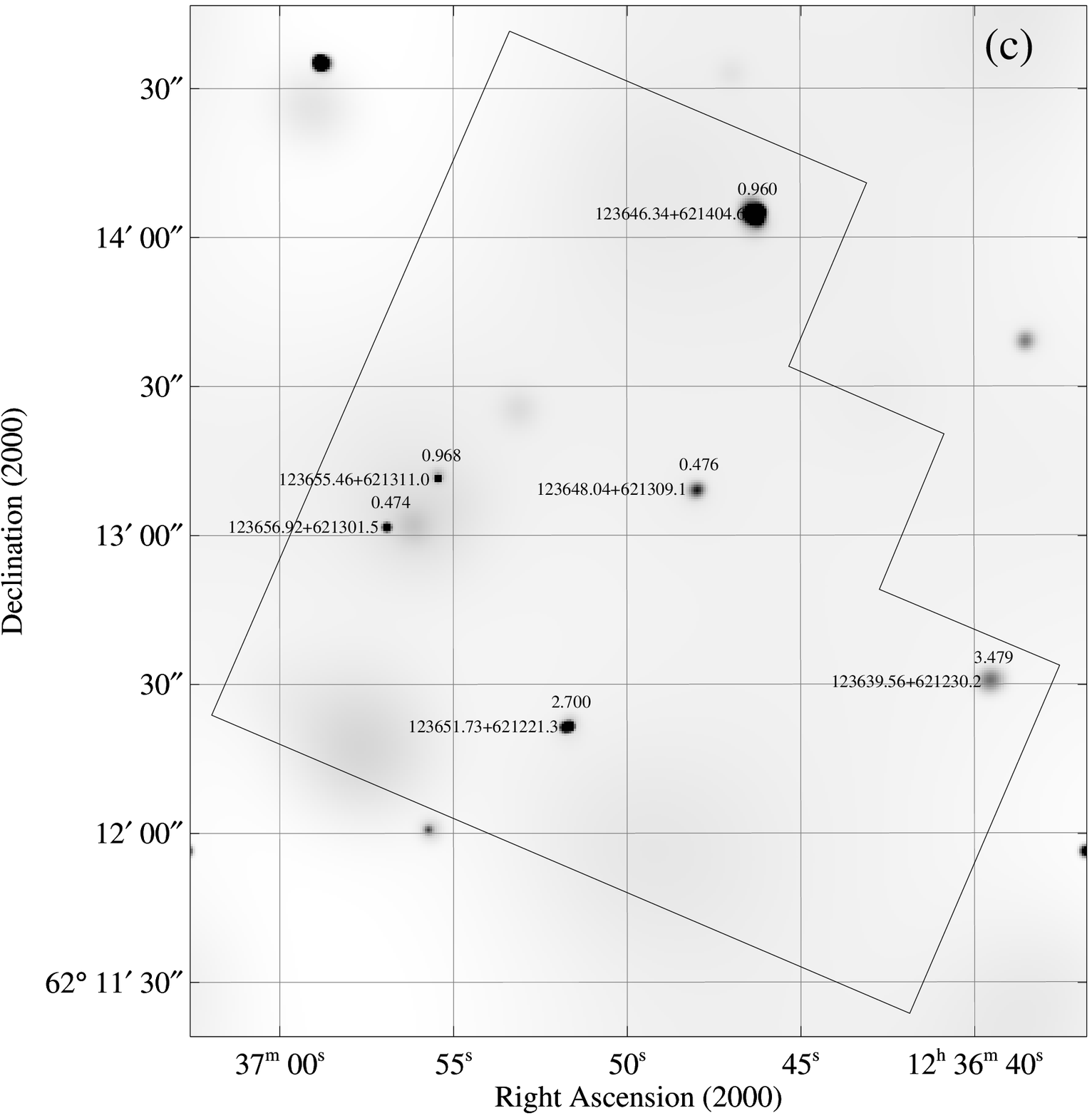}
\end{figure}


\clearpage

\begin{figure}
\vspace{-0.5truein}
\epsscale{0.9}
\figurenum{2}
\plotone{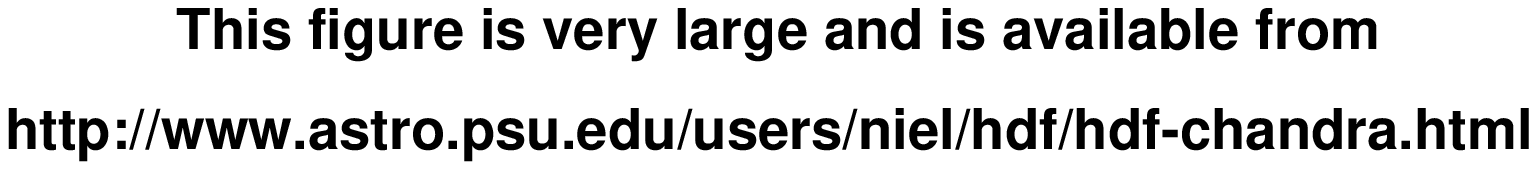}
\caption{
\chandra\ sources detected in the HDF-N circled on the W96 optical image. 
Sources are labeled by their right ascensions.
Green dots are placed above 15 of the 17 optically bright galaxies used in
our stacking analysis (we could not mark two of the optically bright galaxies
near CXOHDFN~J123643.9+621249 and CXOHDFN~J123649.7+621312 due to
symbol crowding). 
\label{optical}}
\end{figure}

\clearpage


\begin{figure}
\vspace{-0.5truein}
\epsscale{0.9}
\figurenum{3}
\plotone{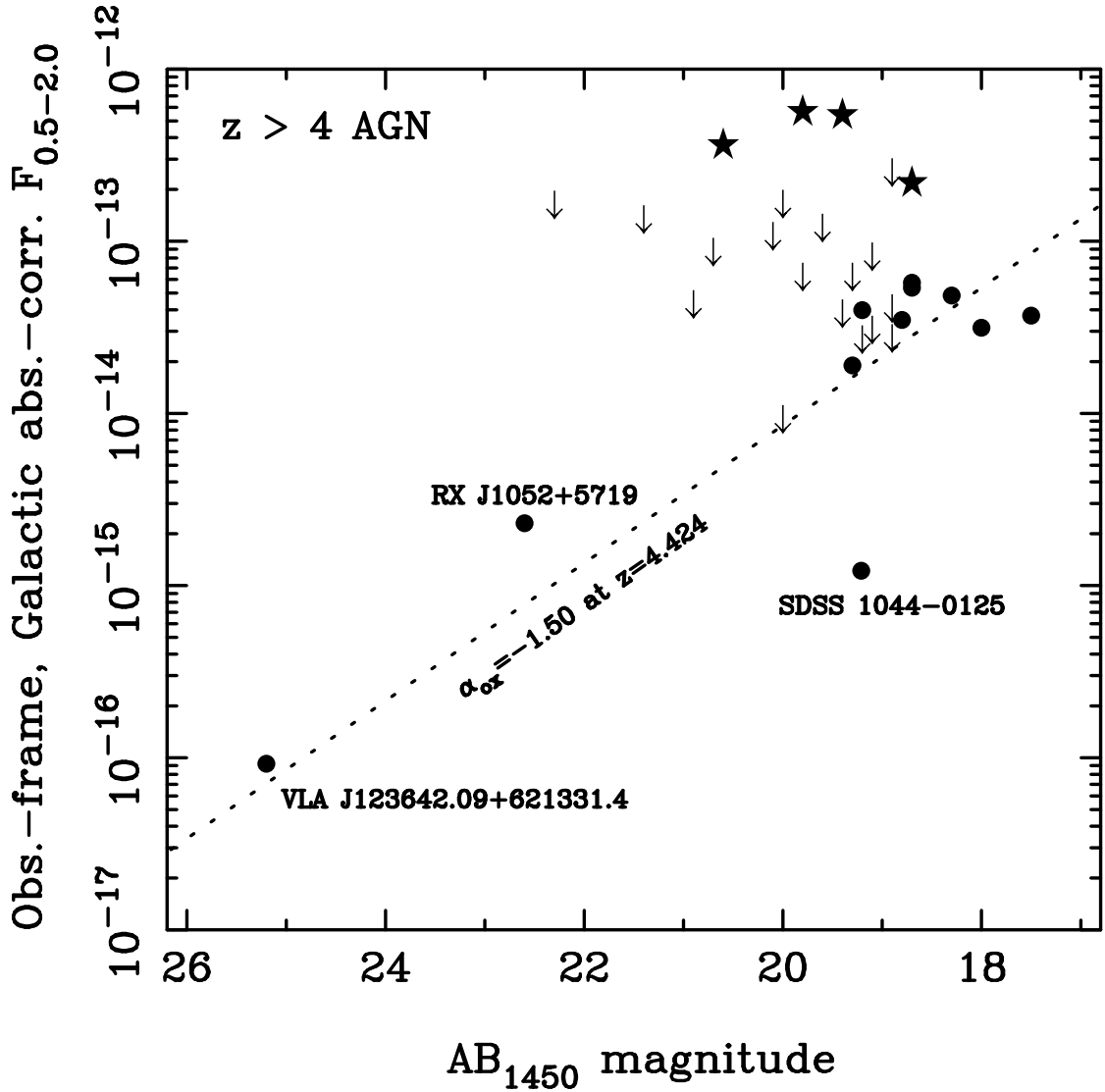}

\caption{
AB$_{1450}$ magnitude versus observed-frame, Galactic absorption-corrected 
0.5--2.0~keV flux for AGN at $z>4$ (adapted from Kaspi et~al. 2000 and
Brandt et~al. 2001). The solid dots are 
X-ray detected radio-quiet AGN, the stars are X-ray detected 
radio-loud blazars, and the arrows are X-ray upper limits.
The oblique dotted line shows the $\alpha_{\rm ox}=-1.50$ locus for $z=4.424$ 
(assuming an optical continuum slope of \hbox{$\alpha_{\rm o}=-0.79$} and an 
X-ray power-law photon index of $\Gamma=2$). 
We have labeled \vla; provided its likely redshift of $z=4.424$ is correct, 
it is both the X-ray faintest and optically faintest $z>4$ AGN detected 
in X-rays. Despite its faintness, its value of \aox\ is similar to that 
of much brighter AGN. 
We have also labeled the notable $z>4$ AGN 
RX~J1052+5719 ($z=4.45$; Schneider et~al. 1998) and
SDSS~1044--0125 ($z=5.80$; Fan et~al. 2000; Brandt et~al. 2001).  
\label{highz}}
\end{figure}


\begin{figure}
\vspace{-0.5truein}
\epsscale{0.9}
\figurenum{4}
\plotone{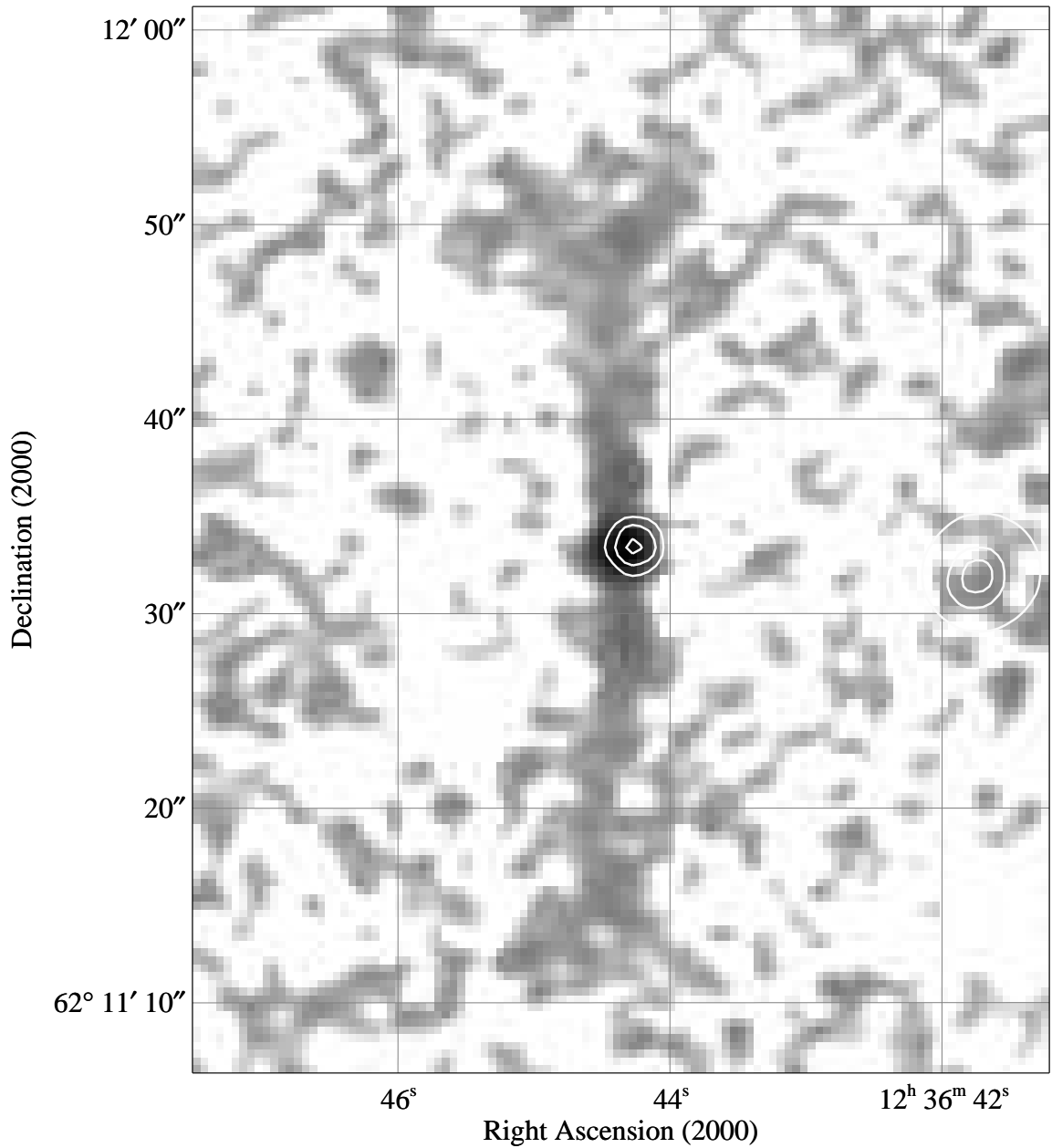}

\caption{Image of the FR~I AGN VLA~J123644.38+621133.0 at 1.4~GHz (Richards 2000) 
with \chandra\ soft-band contours overlaid. The contours are at 16, 32 and
64\% of the maximum pixel value. The X-ray contours near the right-hand
edge of the figure correspond to CXOHDFN~J123641.7+621131. 
\label{fri}}
\end{figure}



\begin{figure}
\vspace{-0.5truein}
\epsscale{0.9}
\figurenum{5}
\plottwo{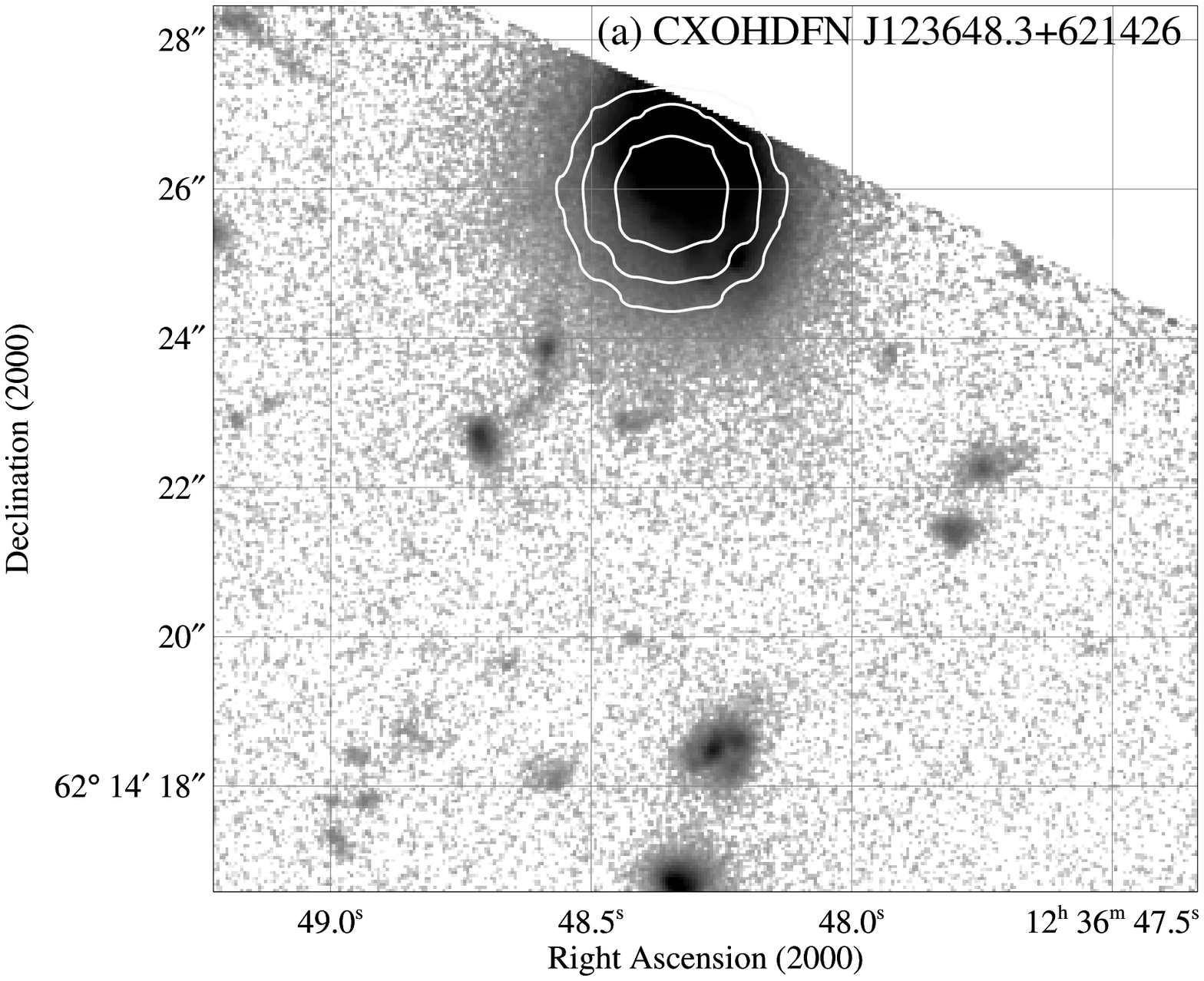}{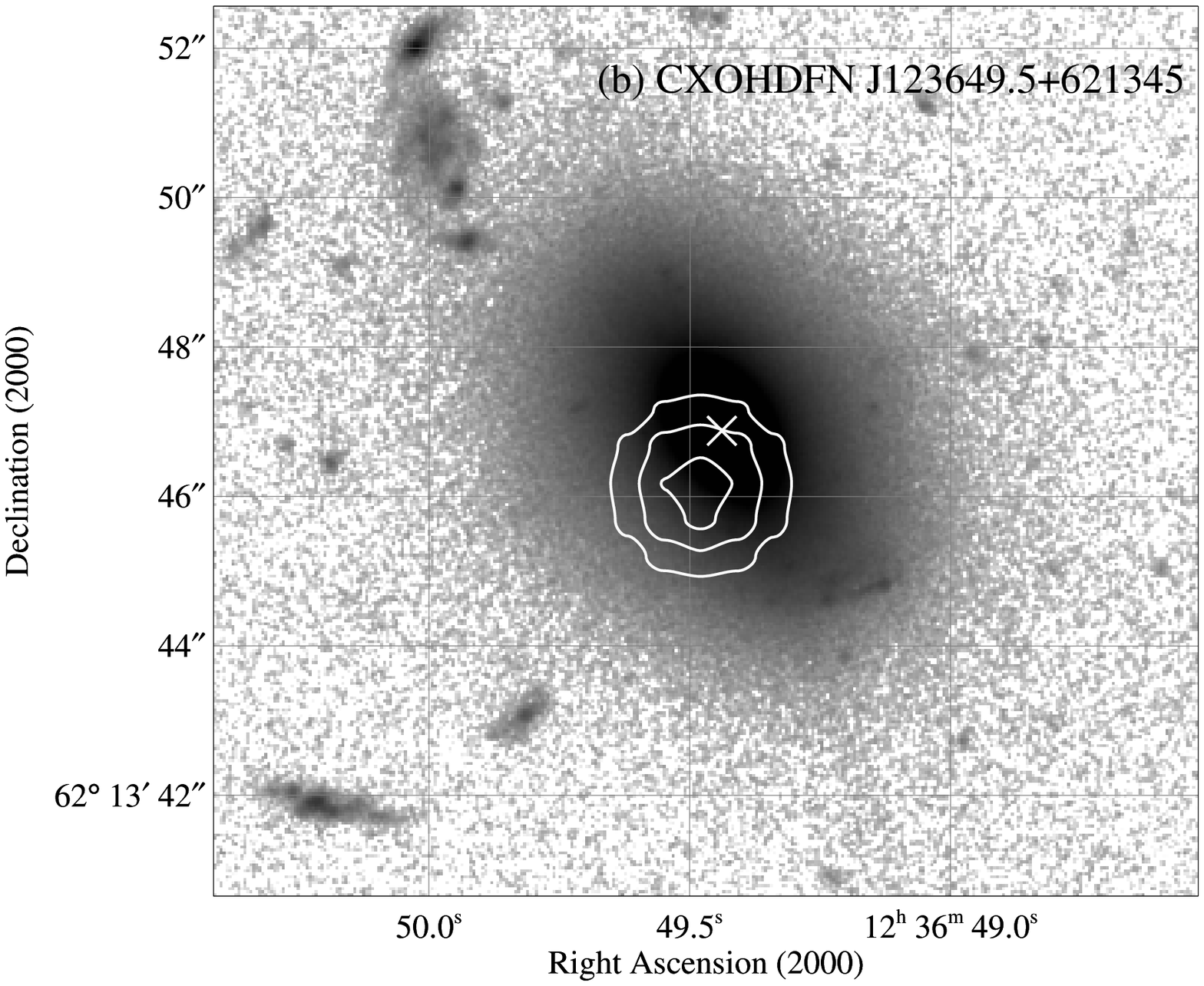}
\plotone{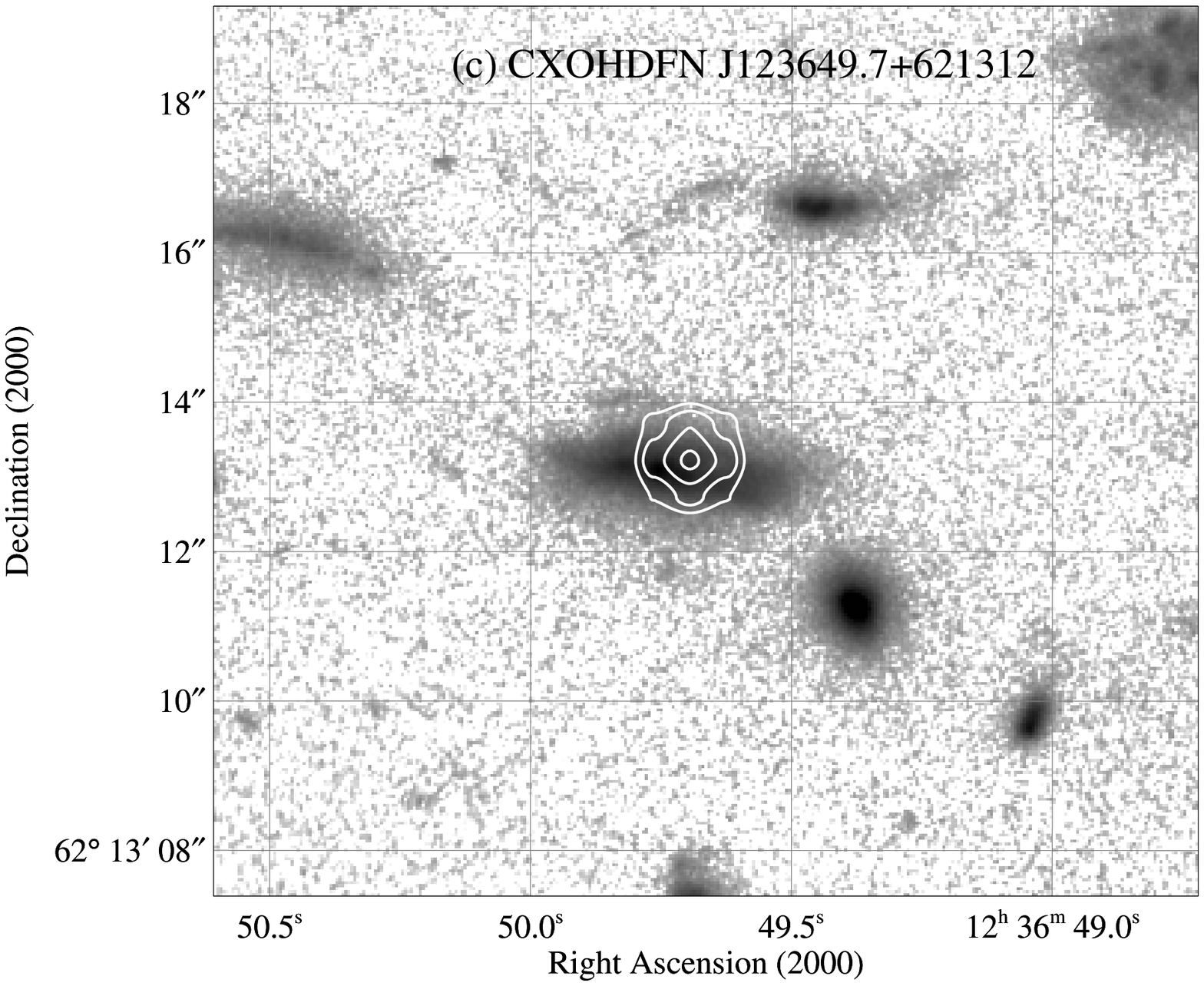}
\caption{$V_{606}$ images from \hst\ with \chandra\ contours for 
(a) CXOHDFN~J123648.3+621426,
(b) CXOHDFN~J123649.5+621345, and 
(c) CXOHDFN~J123649.7+621312
overlaid. Full-band contours are shown for CXOHDFN~J123649.5+621345;
soft-band contours are shown for the other two sources. 
CXOHDFN~J123648.3+621426 lies near the edge of the HDF-N image. 
Note that the X-ray centroid for CXOHDFN~J123649.5+621345 is 
offset from the nucleus of the bright elliptical (marked with
a cross). 
\label{opticalcutouts}}
\end{figure}




\begin{figure}
\vspace{-0.5truein}
\epsscale{0.7}
\figurenum{6}
\plotone{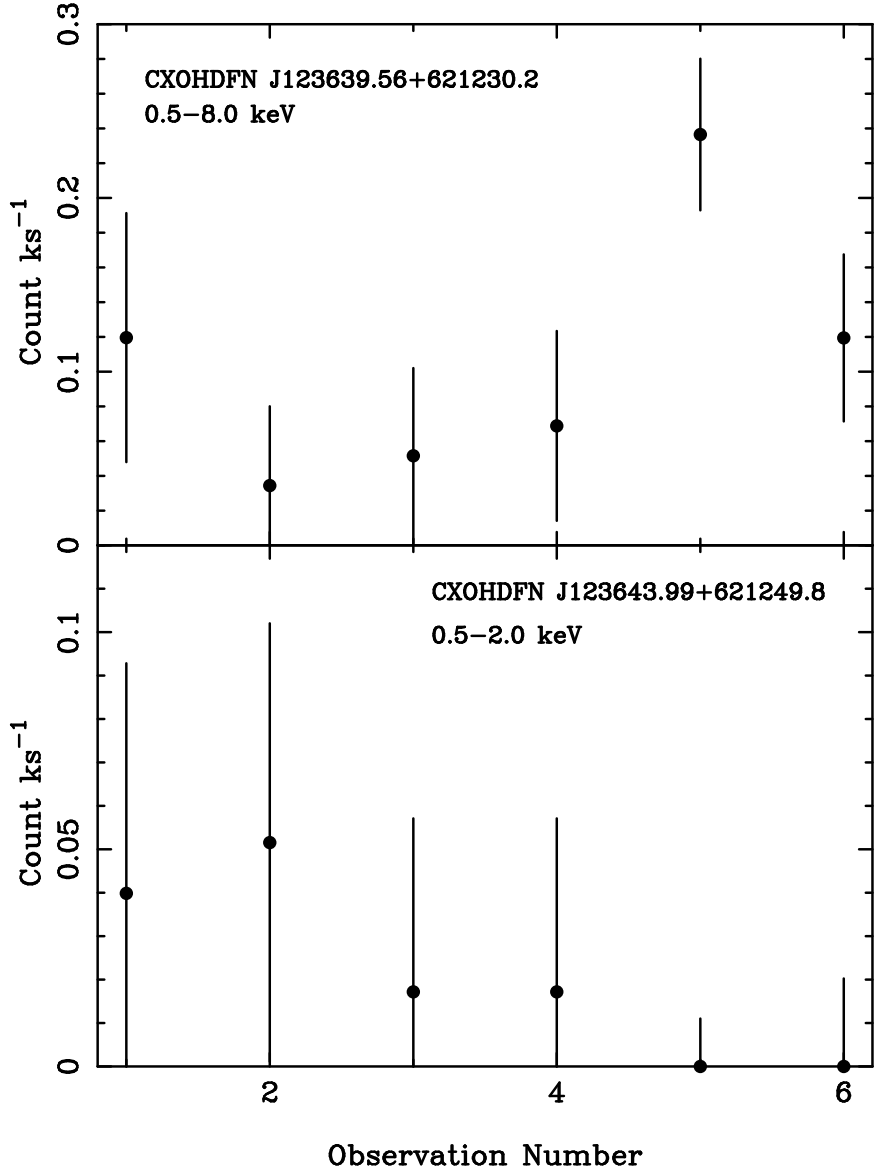}
\caption{\chandra\ light curves for 
CXOHDFN~J123639.5+621230 and CXOHDFN~123643.9+621249 in the
full band and soft band, respectively. The mean count rate is shown
for each of the six observations comprising the \chandra\ data set;
the first four observations are those listed in Table~1 of H01, 
and the last two are those described in \S2. The exposure times
vary substantially from observation to observation, and the 
time gaps between the observations vary. Error bars 
have been calculated following Gehrels (1986). 
\label{variability}}
\end{figure}


\begin{figure}
\vspace{-0.5truein}
\epsscale{1.0}
\figurenum{7}
\plotone{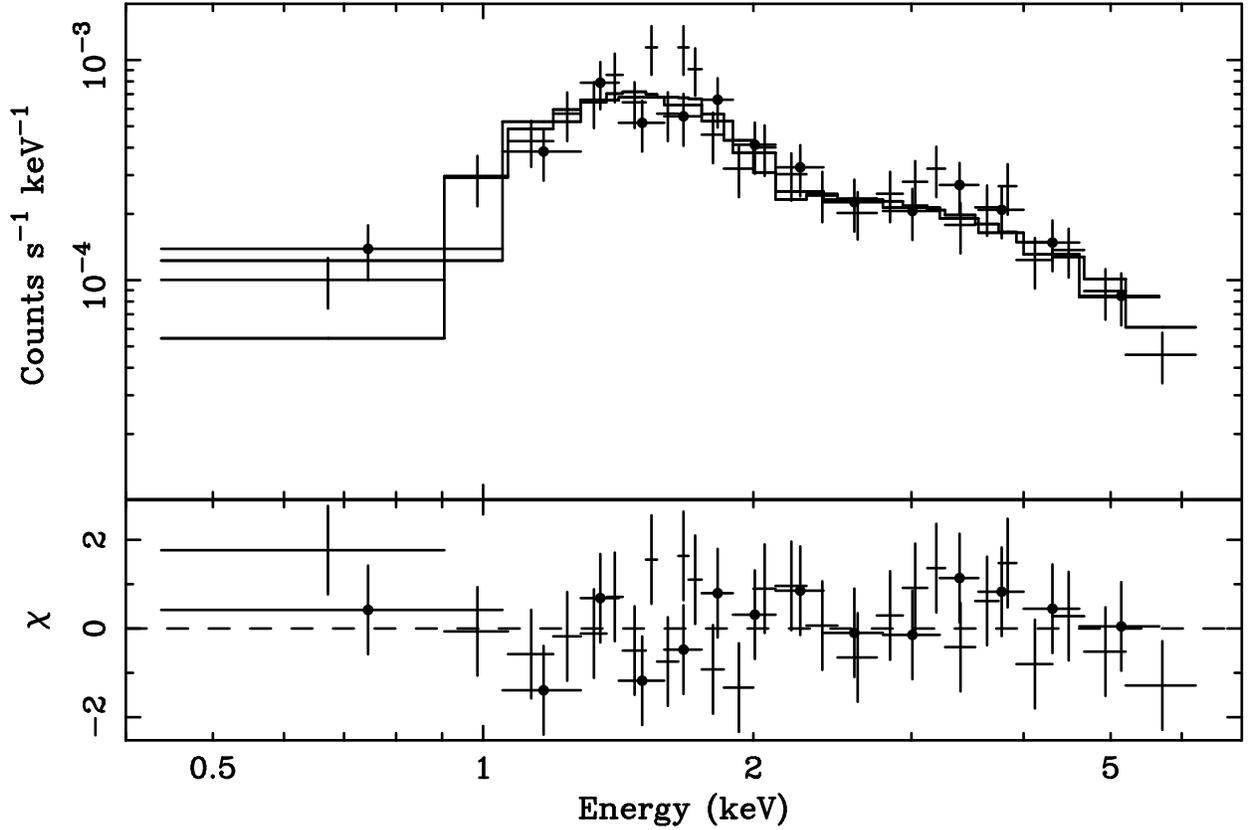}
\caption{\chandra\ ACIS-I observed-frame spectra of CXOHDFN~J123646.3+621404  
along with the best-fitting absorbed power-law model; the best-fitting power-law
photon index is $\Gamma=1.57^{+0.26}_{-0.22}$, and the intrinsic
column density is $(3.97^{+1.41}_{-0.94})\times 10^{22}$~cm$^{-2}$.
The solid dots show the data taken when the focal 
plane temperature was $-110^{\circ}$~C, and the plain crosses
show the data taken when the focal plane temperature was $-120^{\circ}$~C. The ordinate
for the lower panel, labeled $\chi$, shows the fit residuals in terms of
standard deviation with error bars of size one. 
\label{chandraspectrum}}
\end{figure}


\begin{figure}
\vspace{-0.5truein}
\epsscale{1.0}
\figurenum{8}
\plotone{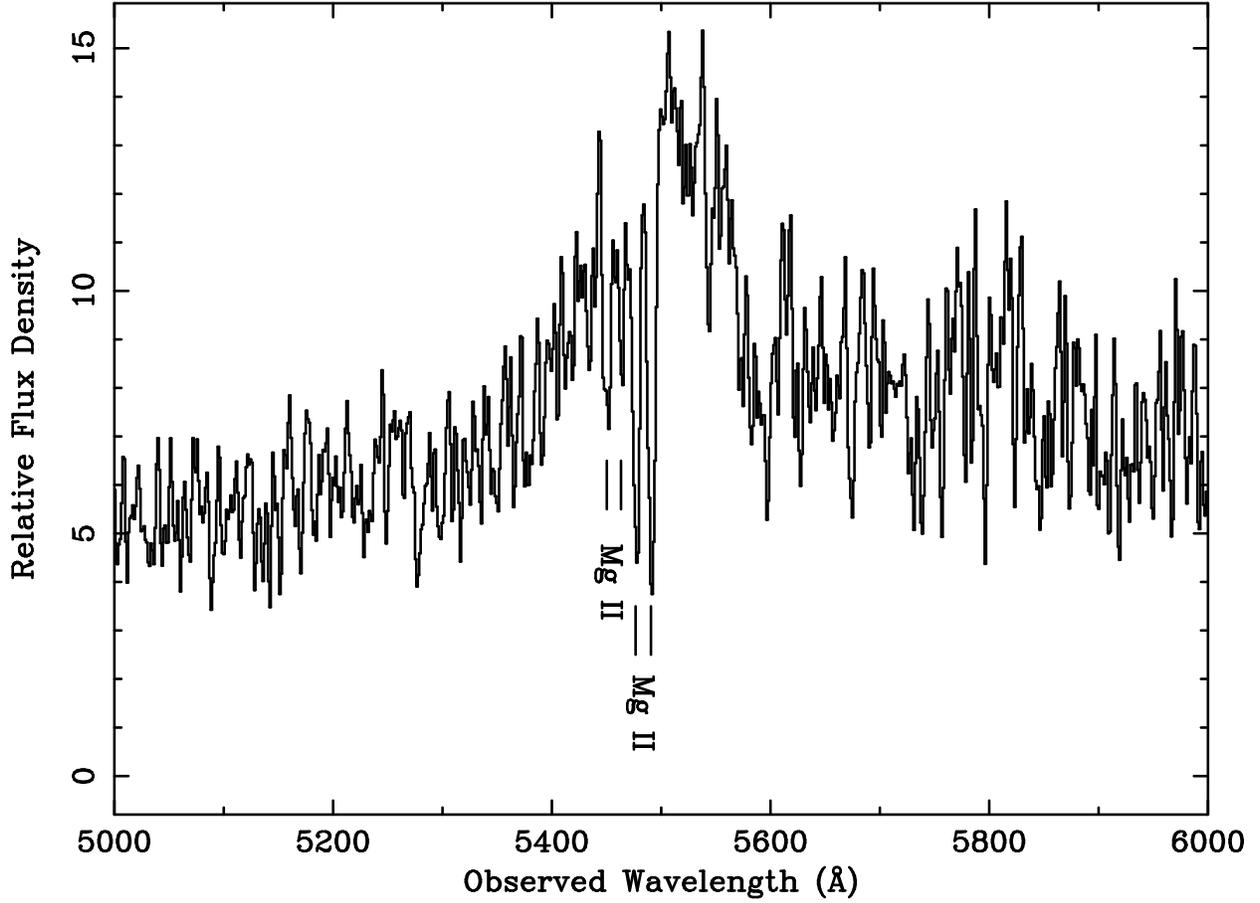}
\caption{Keck Low-Resolution Imaging Spectrograph (LRIS; Oke et~al. 1995) spectrum
of the $z=0.960$ AGN associated with CXOHDFN~J123646.3+621404. The spectral 
resolution is $\approx 3$~\AA. This spectrum was taken by the DEEP collaboration 
(see Phillips et~al. 1997), and only the region near the Mg~{\sc ii} emission line 
is shown. Note the two blueshifted Mg~{\sc ii} absorption doublets superimposed upon
the broad Mg~{\sc ii} emission line. 
\label{keckspectrum}}
\end{figure}


\begin{figure}
\vspace{-0.5truein}
\epsscale{0.8}
\figurenum{9}
\plotone{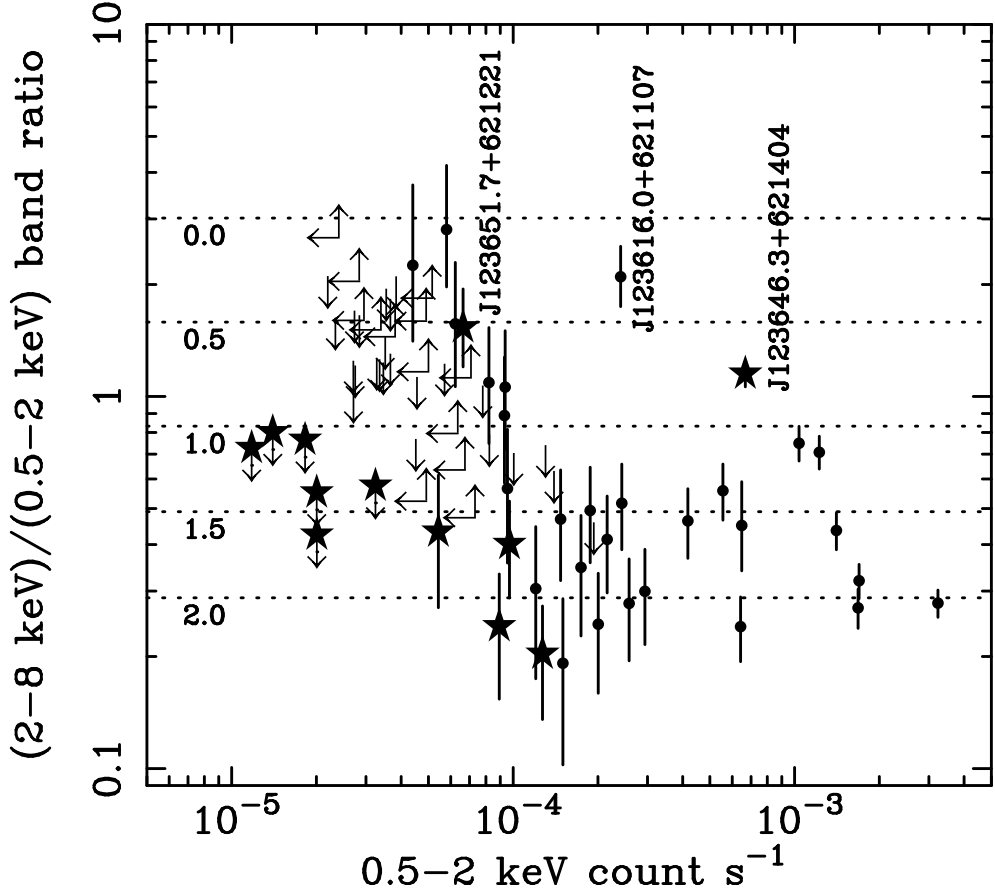}
\caption{Band ratio as a function of soft-band (0.5--2.0~keV) count rate for 
\chandra\ sources. Plotted are the 479.7~ks \chandra\ sources in 
Table~2 (stars) as well as the other sources studied by H01 with 221.9~ks
(solid dots and plain arrows for limits). Dotted lines are labeled with 
the photon indices which correspond to a given band ratio (assuming only
Galactic absorption). Note that the
faintest HDF-N sources occupy a region at low count rate and relatively 
low band ratio that has not, until now, been seen for individual sources. 
Only one source presented in this paper is not plotted here; this
source, CXOHDFN~J123649.5+621345, was only detected in the 
full band. 
\label{bandratio}}
\end{figure}


\begin{figure}
\vspace{-0.5truein}
\epsscale{0.8}
\figurenum{10}
\plotone{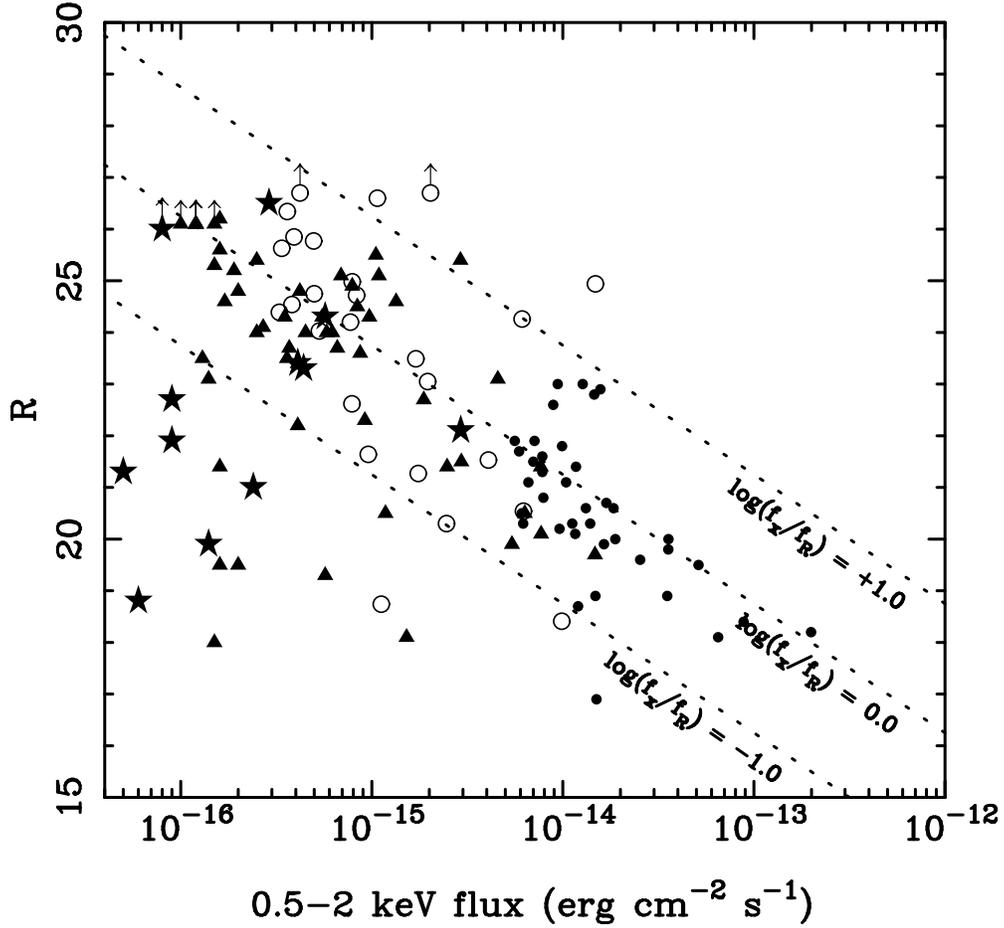}
\caption{Plot of $R$ magnitude versus 0.5--2.0~keV flux for X-ray sources. Plotted
are the 12 \chandra\ HDF-N area sources in Table~2 that are detected in the 
soft band (stars), the 221.9~ks \chandra\ sources from H01 (triangles), the 100~ks 
\chandra\ sources from Mushotzky et~al. (2000; open circles), and the 
\rosat\ sources from Schmidt et~al. (1998; dots). The slanted lines have
been calculated following \S5.1 of H01; they show the typical range of X-ray
to $R$-band flux observed for luminous AGN in the local Universe. 
\label{xrayoptfluxes}}
\end{figure}

\clearpage


\begin{figure}
\vspace{-0.5truein}
\epsscale{0.8}
\figurenum{11previous}
\plottwo{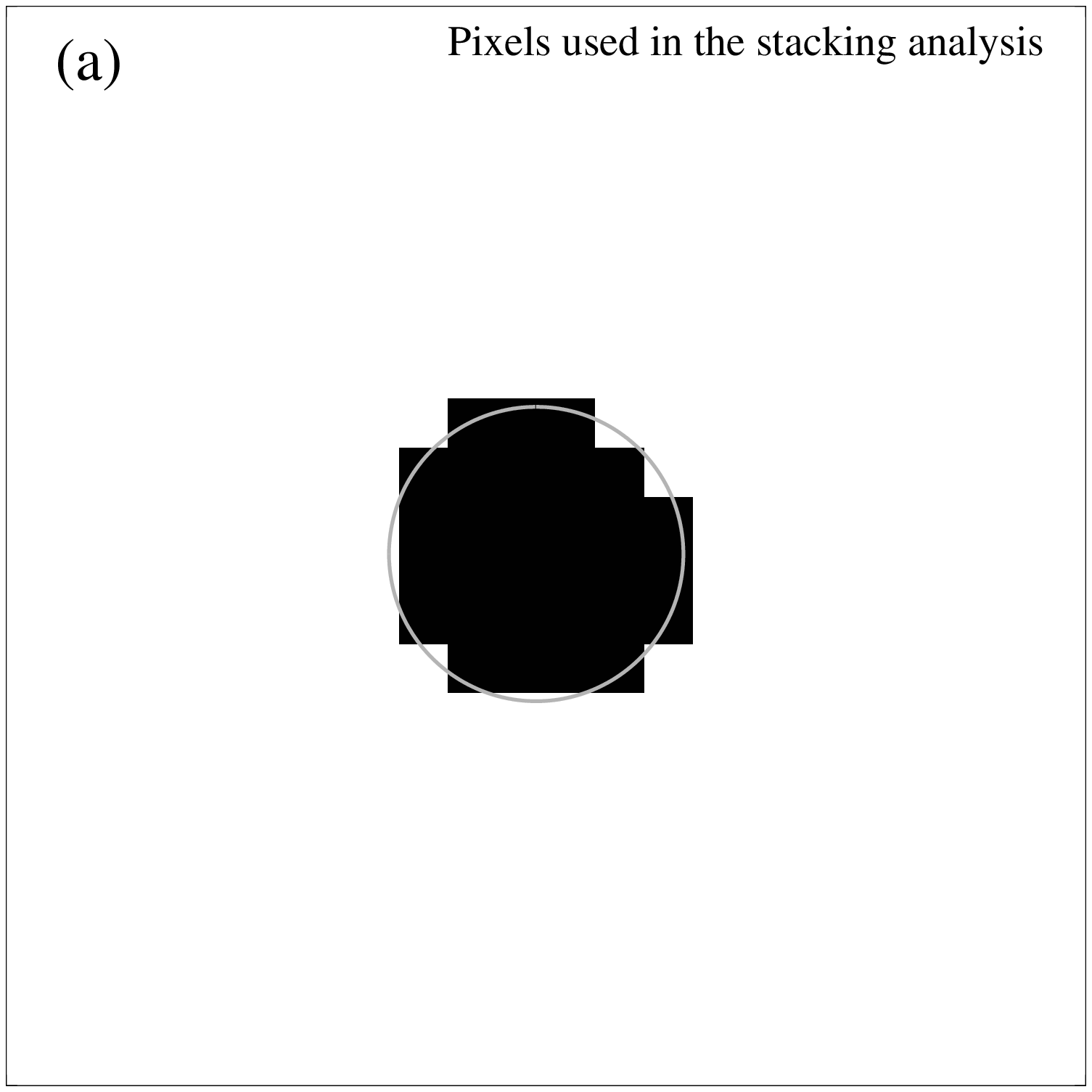}{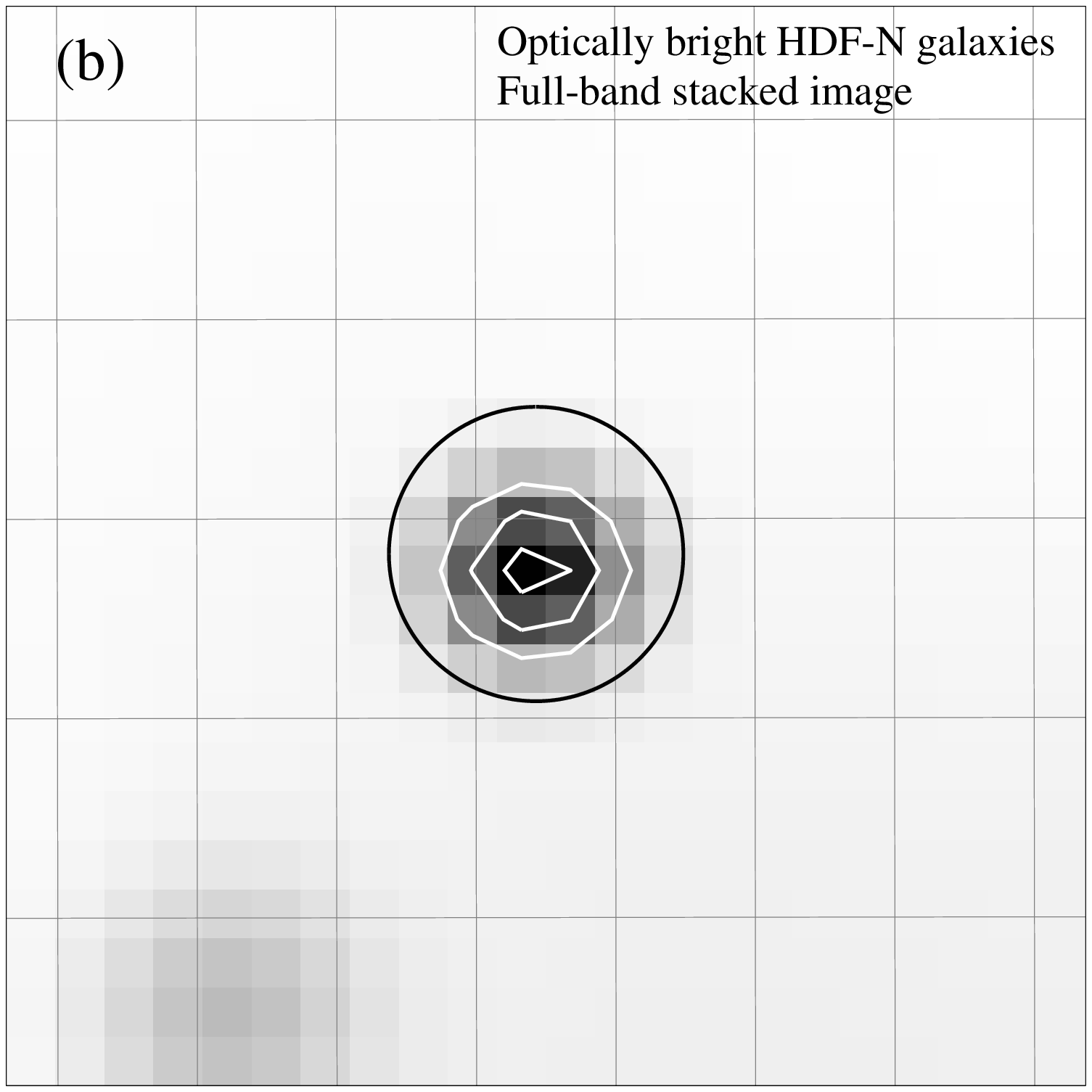} 
\end{figure}
\epsscale{0.8}

\begin{figure}
\vspace{-0.5truein}
\epsscale{0.8}
\figurenum{11}
\plottwo{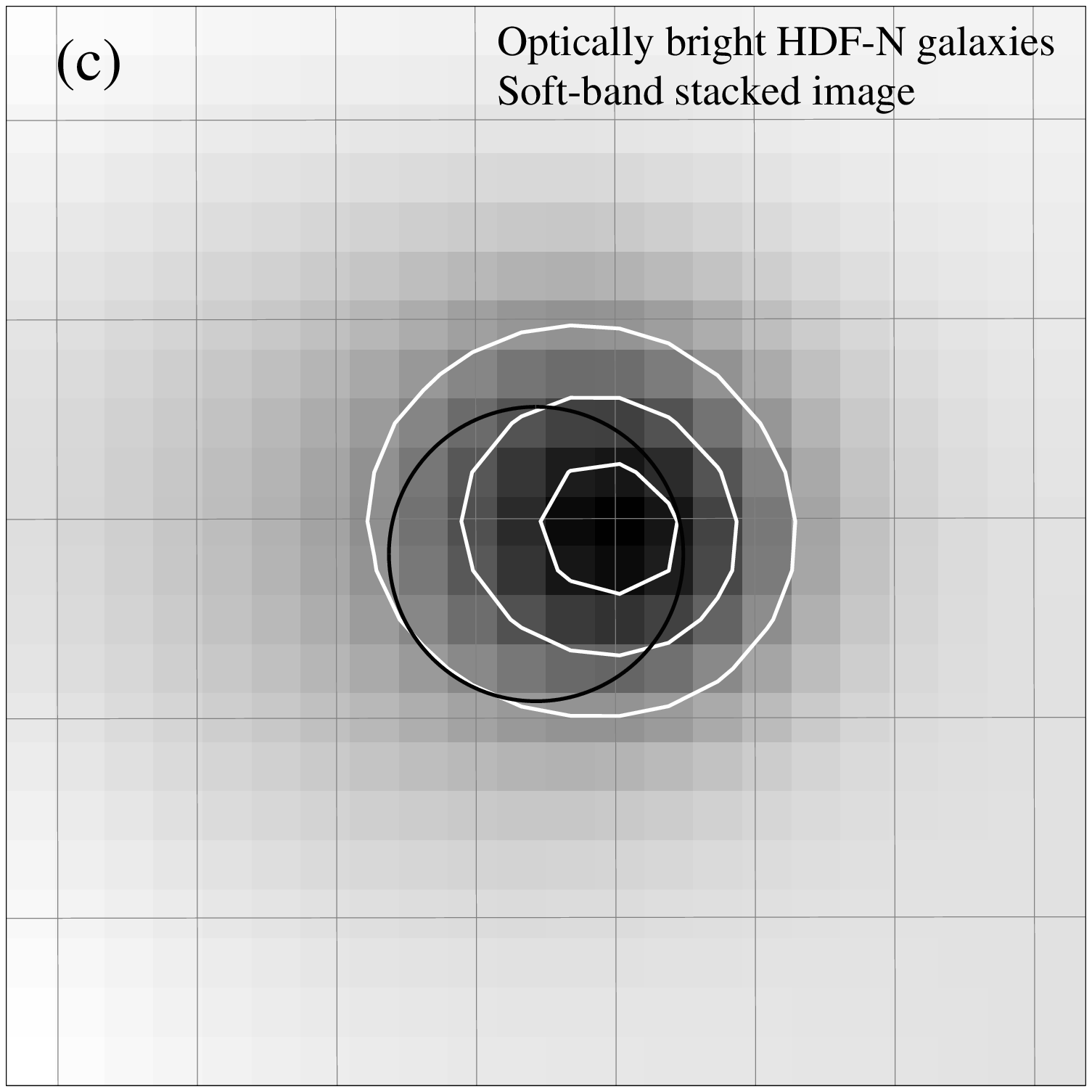}{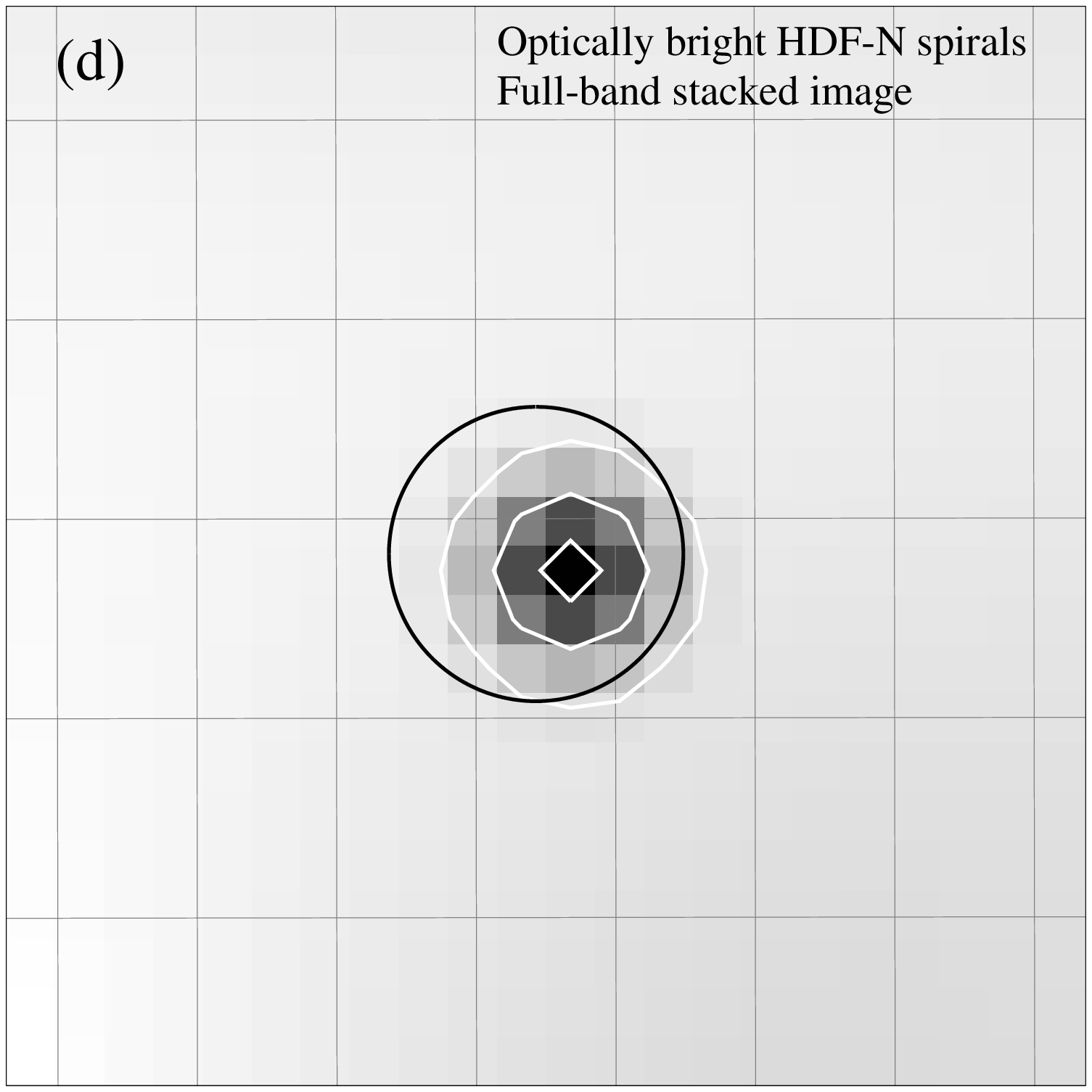}
\caption{Panel (a) shows the pixels used in the stacking analysis (black pixels
were used and white pixels were not).
The other panels show stacked \chandra\ images of 
(b) optically bright HDF-N galaxies in the full band (7.7~Ms effective exposure), 
(c) optically bright HDF-N galaxies in the soft band (7.7~Ms effective exposure), and 
(d) optically bright HDF-N spirals in the full band (5.0~Ms effective exposure). 
The stacked images have been made using the restricted ACIS grade set, and 
they have been adaptively smoothed at the $2.5\sigma$ level using the code of 
Ebeling et~al. (2001). The images are $11^{\prime\prime}\times 11^{\prime\prime}$ 
in size, and each pixel is $0.5^{\prime\prime}$ on a side. In all panels
North is up, and East is to the left. The circle at 
the center of each image is centered on the stacking position
and has a radius of $1.5^{\prime\prime}$. The white contours are drawn at 
75, 85, and 95\% of the maximum pixel value. 
\label{galaxystacking}}
\end{figure}


\begin{figure}
\vspace{-0.5truein}
\epsscale{0.7}
\figurenum{12}
\plotone{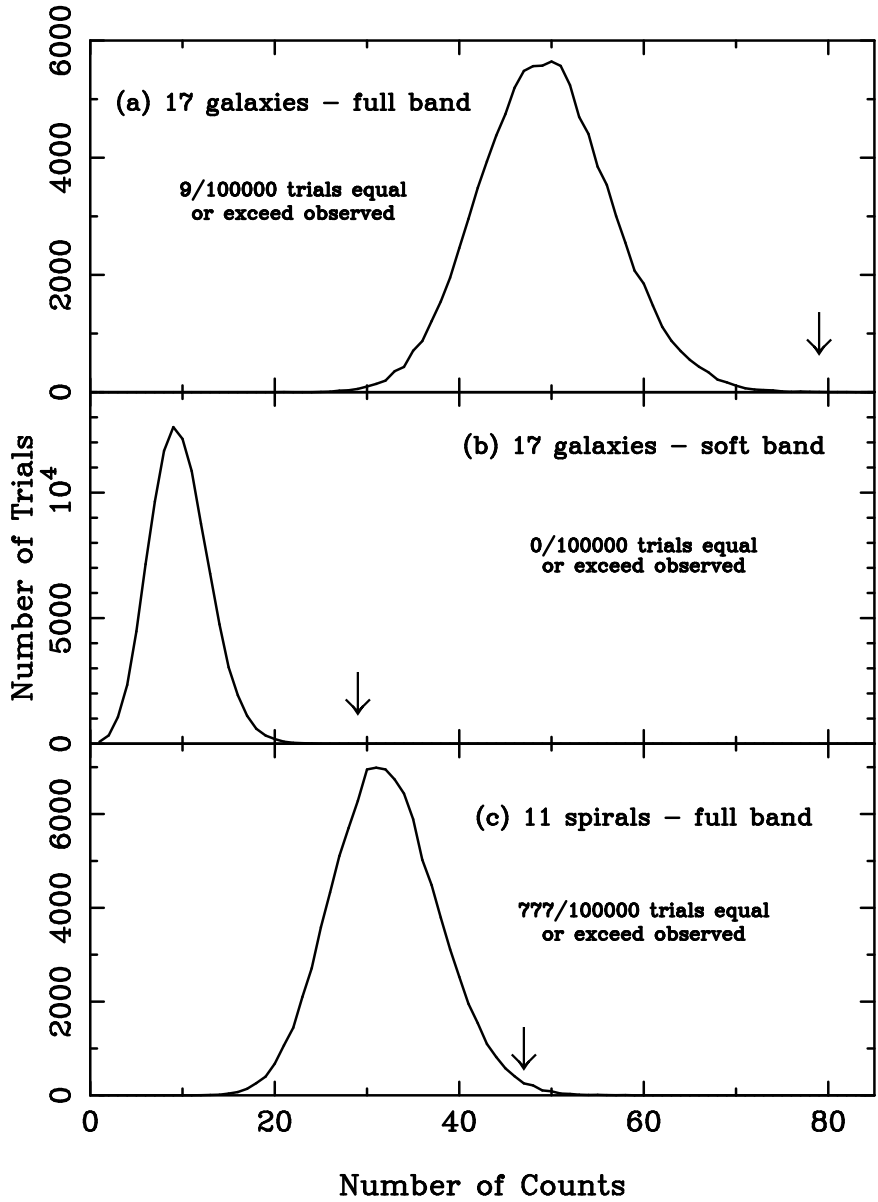}
\caption{Results from our Monte-Carlo testing of the galaxy stacking analysis. 
In each panel, we have performed 100,000 stacking trials at randomly 
selected positions, and we plot the number of trials yielding a given number 
of counts in the 30-pixel source region described in \S3.4.1. Arrows in each
panel show the number of counts actually observed when the stated galaxies
were stacked ($N_{\rm obs}$). Panel (a) is for the full-band stacking of the 17 
galaxies in Table~4, panel (b) is for the soft-band stacking of the 17 galaxies 
in Table~4, and panel (c) is for the full-band stacking of the 11 spiral 
galaxies in Table~4. In each panel we also state the number of random trials
yielding $\geq N_{\rm obs}$ counts.
\label{stackingtests}}
\end{figure}


\end{document}